\documentclass[11pt]{article}  

\usepackage{geometry}
\usepackage[font={small}]{caption}

\usepackage{threeparttable}

\usepackage[utf8]{inputenc}

\usepackage{amsmath,amssymb}

\usepackage{dsfont}
\usepackage{bm}
\usepackage[usenames,dvipsnames]{color}
\usepackage{url}
\usepackage{graphicx}
\usepackage{enumerate}
\usepackage{natbib}
\newcommand{\pkg}[1]{\textsf{#1}}
\newcommand{\bvec}{\left[\begin{array}{c}}
\newcommand{\evec}{\end{array}\right]}
\newcommand{\bmat}[1]{\left[\begin{array}{*{#1}{c}}}
\newcommand{\emat}{\end{array}\right]}

\definecolor{mygray}{gray}{0.5}

\newcommand{\ra}{\rightarrow}

\newcommand{\real}{\mathds{R}}

\newcommand{\df} {\operatorname{df}}
\newcommand{\EV}{\mathds{E}}

\newcommand{\diag} {\operatorname{diag}}
\newcommand{\blockdiag}{\operatorname{blockdiag}}

\newcommand{\argmin}{\operatornamewithlimits{argmin}}

\newcommand{\mba}{\bm{\Phi}}
\newcommand{\ba}{\Phi}

\newcommand{\oper}{\bm{\xi}} 

\newcommand{\mb}{\bm{b}}

\newcommand{\mh}{\bm{h}}
\newcommand{\mH}{\bm{H}}
\newcommand{\mI}{\bm{I}}

\newcommand{\mP}{\bm{P}}

\newcommand{\mgamma}{\bm{\gamma}}

\newcommand{\mtheta}{\bm{\theta}}

\newcommand{\mvartheta}{\bm{\vartheta}}
\newcommand{\mlambda}{\bm{\lambda}}

\begin{document}

\title{Signal Regression Models for Location, Scale and Shape with an Application to Stock Returns}
\author{Sarah Brockhaus$^{1}$\footnote{E-mail for correspondence: \texttt{sarah.brockhaus@stat.uni-muenchen.de}}, Andreas Fuest$^{1}$, Andreas Mayr$^{2}$, and Sonja Greven$^{1}$}
\date{}

\maketitle

\begin{center}
\noindent
\small{$^{1}$ Department of Statistics, Ludwig-Maximilians-Universit\"at M\"unchen, Germany \\
$^{2}$ Department of Medical Informatics, Biometry and Epidemiology, Friedrich-Alexander-Universit\"at Erlangen-N\"urnberg, Germany
}
\end{center}

\begin{abstract}
We discuss scalar-on-function regression models where all parameters of the assumed response distribution can be modeled depending on covariates.
We thus combine signal regression models with generalized additive models for location, scale and shape (GAMLSS).
We compare two fundamentally different methods for estimation, a gradient boosting and a penalized likelihood based approach, and address practically important points like identifiability and model choice. Estimation by a component-wise gradient boosting algorithm allows for high dimensional data settings and variable selection. Estimation by a penalized likelihood based approach has the advantage of directly provided statistical inference.
The motivating application is a time series of stock returns where it is of interest to model both the expectation and the variance depending on lagged response values and functional liquidity curves.
\end{abstract}

\textit{Keywords:} Distributional regression; Functional data analysis; Gradient boosting; Penalized maximum likelihood; Scalar-on-function regression; Variable selection. 

\section{Introduction}

The field of functional data analysis \citep{ramsay2005} deals with the special data situation where the observation units are curves. The functions have continuous support, e.g., time, wave-length or space. One typically assumes that the true underlying functions are smooth and theoretically the functions could be measured on arbitrarily fine grids, even though in practice the functions are measured at a finite number of discrete points. Due to technical progress, more and more data sets containing functional variables are available. Regression methods for functional data are of increasing interest and have been developed for functional responses and/or functional covariates, see \cite{morris2015} for a recent review on regression methods with functional data.

In this paper, we consider a case study on returns for the stocks of Commerzbank from November 2008 to December 2010 \citep{fuest2015}. 
The primary interest lies in predicting the variances of the stock returns whereas the modeling of the expectation is only of secondary interest, as it is a well-known empirical fact that the latter is hardly predictable. In contrast, the conditional variance is typically time-varying and strongly serially correlated \cite[see, e.g.,][]{cont2001}. Apart from the returns, however, our data set includes rich information on the market participants’ offers and requests as potential predictors, which we use as covariates in the form of liquidity curves. 
Consequently, we want to fit a regression model for the expectation and the variance of a scalar response using several functional and scalar covariates.

So far, most publications dealing with scalar-on-function regression have focused on modeling the conditional expectation. The linear functional model was introduced by \cite{ramsay1991}
\begin{equation}
\label{eq.linfunmodel}
y_i = \beta_0 + \int_{\mathcal{S}} x_i(s)\beta(s)\, \mathrm{d}s + \varepsilon_i,
\end{equation}
with continuous response $y_i$, $i=1,\ldots,N$, functional covariate $x_i(s)$, $s \in \mathcal{S}$, with $\mathcal{S}$ a closed interval in $\real$, intercept $\beta_0$, functional coefficient $\beta(s)$ and errors $\varepsilon_i \overset{iid}{\sim} N(0,\ \sigma^2)$. Many extensions of this model have been proposed, concerning the response distribution, non-linearity of the effect and inclusion of further covariate effects. For non-normal responses, generalized linear models \citep[GLMs,][]{nelder1972} with linear effects of functional covariates
\citep[e.g.,][]{marx1999,ramsay2005,muller2005gener,goldsmith2011} 
and generalized additive models \citep[GAMs,][]{hastie1986} 
with non-linear effects of functional covariates \citep[e.g.,][]{james2005,mclean2014} have been discussed.
As distribution free approach quantile regression \citep{koenker2005} with functional covariates has been considered, see, e.g., \cite{ferraty2005,cardot2005,chen2012}.
The most important differences between those models are the choice of the basis representation for the functional covariate and/or the functional coefficient, and the choice of the fitting algorithm. Common choices for the bases are functional principal components and splines. The estimation is mostly done by (penalized) maximum-likelihood approaches.
However, functional regression models that model simultaneously several parameters of the conditional response distribution have not been considered.     
This can be overcome in the framework of generalized additive models for location, scale and shape (GAMLSS), introduced by \cite{rigby2005}, where all distribution parameters can be modeled depending on covariates. For estimation, Newton-Raphson or Fisher-Scoring is used to maximize the (penalized) likelihood. \cite{mayr2012} estimate GAMLSS in high-dimensional data settings by a component-wise gradient boosting algorithm. \cite{klein2015} discuss Bayesian inference for GAMLSS and denominate the model as structured additive distributional regression, as the modeled distribution parameters are not necessarily location, scale and shape but rather determine those characteristics indirectly. Inference is based on a Markov chain Monte Carlo simulation algorithm with distribution-specific iteratively weighted least squares approximations for the full conditionals.
For each of these three approaches, complex parametric distributions like the (inverse) Gaussian, Weibull or Negative Binomial distribution can be assumed for the response variable.
None of these three papers discusses the incorporation of functional covariates into  GAMLSS.

\cite{wood2015} propose an estimation framework for smooth models with (non-)expo\-nential families, including certain GAMLSS as special case. In the current implementation three response distributions for GAMLSS - a normal location scale, a two-stage zero-inflated Poisson and a multinomial logistic model - are available. The focus lies on stable smoothing parameter estimation and model selection.
Already in an earlier paper, \cite{wood2011} discusses how to incorporate linear functional effects into GAMs using spline expansions of $\beta(s)$ and \cite{wood2015} includes an example of a model with ordered categorical response and one functional covariate. The functional model terms can be included in the mentioned GAMLSS within the implementation in \textsf{R}-package \pkg{mgcv} \citep{mgcv2016}, using \textsf{R} software for statistical computing \citep{R2015}. Thus, scalar-on-function models for normal location scale, two-stage zero-inflated Poisson and multinomial logistic models are available in the implementation, but have not been discussed so far.   

In this paper, we discuss the extension of scalar-on-function regression models in the spirit of GAMLSS and denominate these models as signal regression models for location, scale and shape. We address practically important points like identifiability and model choice and compare different estimation methods.    
The signal regression terms are specified as in \cite{wood2011} and \cite{brockhaus2015}. The second allows in addition to linear functional effects as in~\eqref{eq.linfunmodel} interaction terms $z_i\int_{\mathcal{S}} x_{ij}(s)\beta_j(s)\, \mathrm{d}s$ or $\int_{\mathcal{S}} x_{ij}(s)\beta_j(s,z_i)\, \mathrm{d}s$ between scalar $z_i$ and functional $x_{ij}(s)$ covariates.
The functional coefficients $\beta_j(s)$ can be expanded in a spline basis, such as P-splines \citep{eilers1996}, or in the basis spanned by the functional principal components \citep[FPCs, see, e.g.,][chap.~8]{ramsay2005} of the functional covariate. 
Representing the functional covariate and coefficient in FPCs, the scores of the FPCs are used like scalar covariates, see Section~\ref{sec.specEffectsSignal} where we will compare and discuss different choices of basis functions.
We propose and compare two fundamentally different estimation approaches, based on gradient boosting \citep{mayr2012} and (penalized) maximum likelihood \citep{rigby2005,wood2015}.
No estimation method is generally superior, as each is suited to particular situations, as will be discussed in this paper. Boosting allows for high dimensional data settings with more covariate effects than observations and variable selection, but does not provide direct statistical inference. The maximum likelihood approaches imply the usual machinery of statistical inference but are not applicable in high dimensional data settings.

The combination of scalar-on-function regression and GAMLSS is motivated by the application on stock returns but covers a much broader range of response distributions and functional effects. The stock returns are a continuous real-valued response and we will use normal and $t$ location scale models. We model the linear functional effects for the liquidity curves using P-splines or FPC bases.

The remainder of the paper is structured as follows: in Section~\ref{sec.genModel} we formulate the GAMLSS including functional effects. In Section~\ref{sec.specEffects} we give details on possible covariate effect terms, focusing on effects of functional variables. In Section~\ref{sec.boosting} the gradient boosting approach is presented. In Section~\ref{sec.penloglike} we present estimation by penalized maximum likelihood, commenting on the gamlss-algorithms by~\cite{rigby2005,rigby2014} and the smooth regression models by~\cite{wood2015}. A comparison between the estimation methods highlighting pros and cons is given in Section~\ref{sec.compare}. In Section~\ref{sec.modelchoice} we comment on criteria for model choice. 
Section~\ref{sec.stocks} shows the analysis of the log-return data assuming a normal or $t$-distribution for the response and using functional liquidity curves as covariates. In Section~\ref{sec.simGAMLSS} we present results of two simulation studies - the first is closely related to the application; the second systematically compares the three discussed estimation algorithms. We conclude in Section~\ref{sec.discussion} with a short discussion and an outlook on future research.
In the appendix we give details on the implementation of the methods in \textsf{R} including example code for a small simulated dataset. 
Furthermore, the appendix includes more detailed results of the application and the simulation study.


\section{Generic model}
\label{sec.genModel}
We observe data-pairs from $(Y, X)$
assuming that $Y$ given $X$ follows a conditional distribution $F_{Y|X}$ and $X$ can be fixed or random.
Let the response $Y \in \real$ and scalar covariates in $\real$ and functional covariates in $L^2(\mathcal{S})$, where $L^2$ is the space of square integrable functions on the real interval $\mathcal{S}=[S_1, S_2]$, with $S_1<S_2 \in \real$. If there are several functional covariates, they can have differing domains. One functional covariate is denoted by $X_j(s)$, with $s \in \mathcal{S}$.
We denote the observed data by $(y_i,x_i)$, $i=1,\ldots,N$, and an observed functional covariate by $x_{ij}(s)$, with observation points $s \in (s_1,\ldots,s_R)^\top$, $s_r\in \mathcal{S}$, which are equal for the $i=1,\ldots,N$ observed trajectories. In general, random variables are denoted by upper and realizations by lower case.
To represent a GAMLSS for $Q$~distribution parameters \citep{rigby2005}, the additive regression model has the general form
\begin{equation}
\label{brockhaus.mod}
g^{(q)}(\vartheta^{(q)})=
\boldsymbol{\xi}^{(q)}(Y| X=x) =  h^{(q)}(x) = \sum_{j=1}^{J^{(q)}} h^{(q)}_j(x),~~ q=1, \ldots, Q,
\end{equation}
where $\vartheta^{(q)}$ is the $q$th distribution parameter of the response distribution and $g^{(q)}$ is the corresponding link function relating the distribution parameter to its linear predictor $h^{(q)}(x)$. Equivalently, we use a transformation function $\boldsymbol{\xi}^{(q)}$, which is the composition of the function yielding the $q$th parameter of the conditional response distribution and the link function $g^{(q)}$. To represent a GLM the model contains only one equation, $Q=1$, for the expectation, $\vartheta^{(1)}=\mu$, and the transformation function is the composition of the expectation $\mathbb{E}$ and the link function $g^{(1)}$, i.e.~$\boldsymbol{\xi}^{(1)} = g^{(1)} \circ \mathbb{E}$.
More generally, each function in the vector of transformation functions is the composition of a parameter function and a link function. For example, for normally distributed response, the transformation functions can be the expectation composed with the identity link, $\boldsymbol{\xi}^{(1)}=\mathbb{E}$, and the variance composed with the log link, $\boldsymbol{\xi}^{(2)}= \log \circ \mathbb{V}.$ The framework allows for many different response distributions, including e.g.~the (inverse) normal, $t$-, and gamma distribution. See \cite{rigby2005} for an extensive list of response distributions.   

The linear predictor for the $q$th parameter, $h^{(q)}$, is the additive composition of  covariate effects $h^{(q)}_j$. Each effect $h^{(q)}_j(x)$ can depend on one covariate for simple effects or on a subset of covariates in~$x$ to form interaction effects. Linear, non-linear and interaction effects of scalar covariates are possible in the GAMLSS proposed by \cite{rigby2005}, \cite{mayr2012}, \cite{klein2015} and \cite{wood2015}. The latter allows additionally for linear functional effects, but is in the current implementation restricted to three distribution families.
We extend the existing models by allowing for linear functional effects and interaction terms between scalar and functional covariates for general response distributions. 
Table~\ref{tab.effects} gives an overview on possible covariate effects.  
In the following section, we will discuss the setup for the effects, with a special emphasis on effects of functional covariates. 
\begin{table}[ht]
\caption{\label{tab.effects}Covariate effects: Overview of possible effects that can be specified in the linear predictors. }
\centering
\begin{tabular}{lll}
covariate(s) & type of effect & $h_j^{(q)}(x)$  \\
\hline
(none) &  intercept & $\beta_0$  \\
scalar variable $z_1$ & linear effect & $z_1 \beta$  \\
& smooth effect & $f(z_1)$ \\
\hspace*{3mm} plus scalar $z_2$ & linear interaction & $z_1 z_2 \beta$  \\
& smooth interaction & $f(z_1, z_2)$  \\
\hline
grouping variable $g$ & group-specific intercept & $\beta_g$ \\
\hspace*{3mm} plus scalar $z$ & group-specific linear effect & $z \beta_g$  \\
\hline
functional variable $x(s)$ & linear functional effect & $\int_{\mathcal{S}} x(s) \beta(s)\, \mathrm{d}s$ \\
\hspace*{3mm} plus scalar $z$ & linear interaction & $z \int_{\mathcal{S}} x(s) \beta(s)\, \mathrm{d}s$ \\
                              & smooth interaction & $\int_{\mathcal{S}} x(s) \beta(z,s)\, \mathrm{d}s$ \\
\end{tabular}
\end{table}
%


\section{Specification of effects}
\label{sec.specEffects}
We represent smooth effects by basis expansions, representing each partial effect, $h_j^{(q)}(x)$, as linear term
\begin{equation}
\label{eq.hjk}
h_j^{(q)}(x) = \mb_j^{(q)}(x)^\top \mtheta_j^{(q)},
\end{equation}
where $\mb_j^{(q)}: \mathcal{X} \rightarrow \real^{K_j^{(q)}}$, with $\mathcal{X}$ being the space of the covariates $X$, is a vector of basis  evaluations depending on one or several covariates and $\mtheta_j^{(q)}$ is the vector of basis coefficients that is to be estimated. The effects are regularized by a quadratic penalty term,
\begin{equation}
\label{eq.pen}
 \mtheta_j^{(q)\top} \mP_j^{(q)}(\mlambda) \mtheta_j^{(q)},
\end{equation}
which for most covariate effects is taken as $\mP_j^{(q)}(\mlambda)=\lambda_j^{(q)}\mP_j^{(q)}$, with fixed penalty matrix $\mP_j^{(q)}$ and smoothing parameter $\lambda_j^{(q)} \geq 0$ from the vector of all smoothing parameters $\mlambda$. 
Thus, the design matrix for each effect contains rows of basis evaluations $\mb_j^{(q)}(x_i)^\top$, $i=1,\ldots,N$, and the corresponding coefficients are regularized using the penalty matrix $\mP_j^{(q)}(\mlambda)$.
For better readability we skip the superscript $(q)$ in the following description of covariate effects.
\\\\
For the effects of scalar covariates we refer to \cite{rigby2005}, \cite{mayr2012} and \cite{wood2015}, where even more effects of scalar variables than those listed in Table~\ref{tab.effects} are described. The specification of effects of functional covariates is discussed in the following two subsections.


\subsection{Signal regression terms}
\label{sec.specEffectsSignal}
Let a functional covariate $x_j(s)$ with domain $\mathcal{S}$ be observed on a grid $(s_1, \ldots, s_R)^\top$.
For a linear functional effect $\int_{\mathcal{S}} x_j(s) \beta_j(s)\, \mathrm{d}s$, also called signal regression term, the integral is approximated using weights $\Delta(s)$ from a numerical integration scheme \citep{wood2011} giving
\begin{align}
\label{eq.funCov}
\mb_j (x_j(s) )^\top
 &=
[  \tilde{x}_j(s_1)\  \cdots\  \tilde{x}_j(s_R)  ]
\left[ \mba_j(s_1)\ \cdots \ \mba_j(s_R)
\right]^\top \\ \nonumber
&= \left[\sum_{r=1}^R \tilde{x}_j(s_r) \ba_1(s_r)\ \cdots\ \sum_{r=1}^R \tilde{x}_j(s_r) \ba_{K_j}(s_r)\right],
\end{align}
where $\tilde{x}_j(s)=\Delta(s)x_j(s)$ and $\mba_j(s)$ is a vector of basis functions $\ba_k(s)$, $k=1,\ldots,K_j$, evaluated at~$s$. The penalty matrix~$\mP_j$ is chosen as matching to the basis functions~$\mba_j$, e.g., a squared difference matrix for B-splines \citep{eilers1996}.

To model a linear interaction between a scalar covariate $z$ and a functional covariate, $z\int_{\mathcal{S}} x_j(s) \beta_j(s)\, \mathrm{d}s$, the basis in~\eqref{eq.funCov} is multiplied by~$z$. For a smooth interaction $\int_{\mathcal{S}} x_j(s) \beta_j(z,s)\, \mathrm{d}s$, we use \citep{scheipl2015,brockhaus2015}
\begin{equation}
\label{eq.funCovIntercation}
\mb_j (x_j(s),z )^\top
 = \mb_{j1} (x_{j}(s) )^\top \odot \mb_{j2} (z)^\top,
\end{equation}
where $\odot$ is the row-tensor product, $\mb_{j1}(x_j(s))$ is defined as in~\eqref{eq.funCov} and $\mb_{j2}(z)=\mba_j(z)$ are spline basis-evaluations of the scalar covariate~$z$. A suitable penalty matrix for such an effect can be constructed as \citep[sec.\ 4.1.8]{wood2006}
\begin{equation}
\label{eq.penaltyIntercation}
\mP_{j}(\mlambda) =  \lambda_{j1} \left(\mP_{j1} \otimes \mI_{K_{j2}} \right) +  \lambda_{j2} \left( \mI_{K_{j1}} \otimes \mP_{j2} \right),
\end{equation}
where $\mP_{j1} \in \real^{K_{j1} \times K_{j1}}$, and $\mP_{j2} \in \real^{K_{j2} \times K_{j2}}$ are appropriate penalty matrices for the marginal bases $\mb_{j1}(x_j(s))$ and $\mb_{j2}(z)$, and $\lambda_{j1}$, $\lambda_{j2} \geq 0$ are smoothing parameters.

\subsection{Choice of basis functions and identifiability}
\underline{Spline bases.}
Assuming the coefficient function $\beta_j(s)$ to be smooth, the basis functions $\mba_j$ for the functional linear effect~\eqref{eq.funCov} can be chosen for instance as B-splines, natural splines or thin plate regression splines. Depending on the chosen spline basis a suitable penalty matrix has to be selected. According to the penalty different smoothness assumptions are implied. For example using P-splines \citep{eilers1996}, i.e. B-splines with a squared difference matrix as penalty, it is possible to penalize deviations from the constant line or the straight line using first or second order differences in the penalty. 

Spline based methods require functional observations on dense grids and thus, for sparse grids they have to be imputed in a preprocessing step \citep[see e.g.,][]{goldsmith2011}. The implementations in \pkg{mgcv} and \pkg{FDboost} require functional observations on one common grid.  
\\\\
\underline{Functional principal component basis.}
In functional data analysis, functional principal component analysis \citep[FPCA, see, e.g.,][]{ramsay2005} is a common tool for dimension reduction, and also widely used in functional regression analysis to represent the functional covariates and/or the functional coefficients \citep[cf.,][]{morris2015}.
Let $X_j(s)$ be a zero-mean square-integrable stochastic process and $e_k(s)$ the eigenfunctions of the auto-covariance of $X_j(s)$, with the respective decreasing eigenvalues $\zeta_1 \geq \zeta_2 \geq \cdots \geq 0$. Then $\{e_k(s), k\in \mathbb{N} \}$ form an orthonormal basis for the $L^2(\mathcal{S})$ and the Karhunen-Lo\`eve theorem states that
\[
X_{ij}(s) = \sum_{k=1}^{\infty} Z_{ik} e_k(s),
\]
where $Z_{ik}$ are uncorrelated random variables with mean zero and variance~$\zeta_k$. This means that functional observations $x_{ij}(s)$ can be represented as weighted sums of (estimated) eigenfunctions. The eigenfunctions represent the main modes of variation of the functional variable and are also called functional principal components (FPCs). In practice, the sum is truncated at a certain number of basis functions. For a fixed number of basis functions, the eigenfunctions are the set of orthonormal basis functions that best approximate the functional observations \citep[see, e.g.,][]{ramsay2005}.
Representing both the functional covariate and the functional coefficient by the eigenfunction basis truncated at $K_j$,
\begin{equation}
\int_{\mathcal{S}} x_{ij}(s) \beta_j(s)\, \mathrm{d}s
\approx \sum_{k,l=1}^{K_j} \int_{\mathcal{S}} z_{ik} e_k(s) e_l(s) \theta_l \, \mathrm{d}s
= \sum_{k=1}^{K_j} z_{ik} \theta_k,
\label{eq.fpcaeffect}
\end{equation}
follows from the orthonormality of $e_k(s)$. This approach thus corresponds to a regression onto the estimated first $K_j$ FPC scores~$z_{ik}$ and interaction effects with other scalar covariates can be specified straight-forwardly.
Regularization is usually achieved by using only the first few eigenfunctions explaining a fixed proportion of total variability \citep[cf.,][]{morris2015}, for example 99\%. Additionally, a penalty matrix can be used for regularization. The penalty $\mP_j =\diag(1/\zeta_1,\ldots,1/\zeta_{K_j})$ assumes decreasing and $\mP_j = \diag(1,\ldots,1)$ equal importance of the eigenfunctions.
If the functional covariate is observed on irregular or sparse grids, it is advantageous to use an FPC basis as it can be estimated directly from the data even in this case \citep{yao2005fun}.
Statistical inference is usually done conditional on the eigendecomposition and thus neglects the variability induced by the estimation of the eigenfunctions and FPC scores \citep{goldsmith2013}. 
\\\\
\underline{Implicit assumptions and identifiability.}
Using an FPC basis, $\beta_j(s)$ is assumed to lie in the space spanned by the first $K_j$ eigenfunctions and the estimation depends on the choice of the discrete tuning parameter $K_j$. For the spline representation, $\beta_j(s)$ is assumed to be smooth and to lie within the space spanned by the spline basis. In practice, those assumptions are hard to check.
If almost all variation of the functional covariate can be explained by the first few eigenfunctions, the covariate carries only little information. In this case, identifiability problems can occur for a spline-based approach \citep{scheipl2016} and the estimation of $\beta_j(s)$ might be dominated by the smoothness assumption. When using an FPCA basis, the estimation is dominated by the assumption that $\beta_j(s)$ lies in the span of the first $K_j$ eigenfunctions, with the estimate highly sensitive to the quality of the estimates of $e_k(s)$ as well as to the choice of $K_j$. Higher-order eigenfunctions are usually relatively wiggly and the shape of $\hat{\beta}_j(s)$ can thus change strongly when increasing $K_j$ \citep[see e.g.,][]{crainiceanu2009}.

\section{Estimation by gradient boosting}
\label{sec.boosting}
In this section we discuss the estimation of GAMLSS by a gradient boosting algorithm as introduced by \cite{mayr2012} and implemented in the \textsf{R}~package \pkg{gamboostLSS} \citep{gamboostLSS2015}. We expand the algorithm by effects \eqref{eq.funCov} and \eqref{eq.funCovIntercation} for functional covariates that are available in the \pkg{FDboost}-package \citep{FDboost2016}. 

Gradient boosting is a machine learning algorithm that aims at minimizing an expected loss criterion along the steepest gradient descent \citep{friedman2001}.
The model is represented as the sum of simple (penalized) regression models, which are called base-learners. In our case the base-learners are the models for the effects $h_j^{(q)}(x)$, as defined by~\eqref{eq.hjk} and~\eqref{eq.pen}. We use a component-wise gradient boosting algorithm \citep[see, e.g.,][]{buhlmann2007}, that iteratively fits each base-learner to the negative gradient of the loss and only updates the best-fitting base-learner per step. Thus, models for high dimensional data settings with more covariates than observations can be estimated and variable selection is done inherently, as base-learners that are never selected for the update are excluded from the model. 

Boosting minimizes the expected loss
\begin{equation*}
\hat \mh = \underset{\mh}{ \argmin}\ \EV\ \rho \left( Y, \mh(X) \right),
\end{equation*}
where $\mh(X)=(h^{(1)}(X), \ldots, h^{(Q)}(X))^\top$ and $\rho: \real \times \real^Q \ra [0, \infty)$ is the loss function. 
In practice, for observed data $(y_i, x_i)$, $i=1,\ldots,N$, the theoretical expectation is approximated by the sample mean, yielding optimization of the empirical risk,
\begin{equation}
\hat \mh = \underset{\mh}{ \argmin}\ \frac{1}{N} \sum_{i=1}^N w_i \; \rho \left( y_i, \mh(x_i) \right),
\end{equation}
where $w_i$ are sampling weights that can be used in resampling methods like cross-validation or bootstrapping.
To estimate a GAMLSS via boosting, the negative log-likelihood of the response distribution is used as loss function \citep{mayr2012},
i.e., $\rho \left(y_i, \mh(x_i) \right) = -l(\mvartheta_i , y_i)$, with the log-likelihood $l$ depending on the vector of distribution parameters $\mvartheta_i = (\vartheta_i^{(1)}, \ldots, \vartheta_i^{(Q)})^\top$, with $\vartheta_i^{(q)} = g^{(q)-1}( h^{(q)}(x_i) )$, and the response $y_i$. 
We expand the framework that \cite{mayr2012} developed for boosting GAMLSS by base-learners for signal regression terms, cf.~equation~\eqref{eq.funCov}, and interaction terms between scalar and functional covariates, cf.~equation~\eqref{eq.funCovIntercation}.
In detail, the following boosting algorithm is used: 
\subsubsection*{Algorithm: Gradient boosting for GAMLSS with functional covariates}
\begin{description}
\item[\textit{Step 1:}] Define the bases $\mb_{j}^{(q)}(x,t)$, their penalties $\mP_{j}^{(q)}(\mlambda)$, $j=1,\ldots, J^{(q)}$, $q=1,\ldots,Q$, and the weights~$w_{i}$, $i=1,\ldots, N$.
Select a vector of step-lengths $(\nu^{(1)}, \ldots, \nu^{(Q)})^\top$, with $\nu^{(q)} \in (0,1)$, and a vector of stopping iterations $(m_{\text{stop}}^{(1)},\ldots, m_{\text{stop}}^{(Q)})^\top$.
Initialize the coefficients $\mtheta^{(q)[0]}_j$. \\
Set the number of boosting iterations to zero, $m=0$.
\item[\textit{Step 2:}] (Iterations over the parameters of the response distribution, $q=1, \ldots, Q$)
\begin{enumerate}[(a)]
\item Set $q=1$.
\item If $m > m_{\text{stop}}^{(q)}$ go to step 2(g); \\
otherwise, compute the negative partial gradient of the empirical risk by plugging in the current estimates
$\hat \mh^{[m]}= \left(\hat h^{(1)[m]}, \ldots, \hat h^{(Q)[m]} \right)^\top$,
with $\hat h^{(q)[m]}(x_i) = \sum_j \mb_{j}^{(q)}(x_i) \mtheta_j^{(q)[m]}$,
as
\[ u_{i}^{(q)}= - \left. \frac{\partial}{\partial h^{(q)}}
 \rho \left( y_i, \mh(x_i) \right) \right\vert _{ \mh(x_i) = \hat{\mh}^{[m]}(x_i)}.
\]
\item Fit the base-learners contained in $h^{(q)}$ for $j = 1, \ldots, J^{(q)}$ to $u_{i}^{(q)}$:
\[ \hat{\mgamma}_j = \underset{\mgamma \in \real^{K_j}}{\argmin}\,
 \sum_{i=1}^N w_{i}
\left\{ u_{i}^{(q)} - \mb_{j}^{(q)}(x_i)^\top \mgamma \right\}^2
+ \mgamma^\top \mP_{j}^{(q)}(\mlambda) \mgamma,
\]
with weights $w_{i}$ and penalty matrices $\mP_{j}^{(q)}(\mlambda)$. 
\item Select the best fitting base-learner, defined by the least squares criterion:
\[ j^{\star} = \underset{j=1, \ldots, J^{(q)}}{\argmin}\, \sum_{i=1}^N  w_{i}
\left\{ u_{i}^{(q)} - \mb_{j}^{(q)}(x_i)^\top \hat{\mgamma}_j \right\}^2.
 \]
\item Update the corresponding coefficients of $h^{(q)[m]}$ to
$\mtheta_{j^\star}^{(q)[m]} = \mtheta_{j^\star}^{(q)[m]} + \nu^{(q)} \hat{\mgamma}_{j^\star}$.  
\item Set $\mtheta_j^{(q)[m+1]} = \mtheta_j^{(q)[m]}$, $j = 1, \ldots, J^{(q)}$.  
\item Unless $q=Q$, increase $q$ by one and go back to step 2(b).
\end{enumerate}
\item[\textit{Step 3:}] Unless $m \geq m_{\text{stop}}^{(q)}$ for all $q$, increase $m$ by one and go back to step 2.
\end{description}
Each component of the final model is a linear combination of base-learner fits $\hat{ \oper}^{(q)} (Y_i |  X_i = x_i) = \sum_{j} \hat{h}_j^{(q)[m_{\text{stop}}^{(q)}]}(x_i)$.
In order to get a fair model selection, we specify equal and rather low degrees of freedom for all base-learners by using adequate values for the smoothing parameters $\lambda_j^{(q)}$ \citep{hofner2011}.
Using additionally a small fixed number for the step-length, e.g., $\nu^{(q)}=0.1$, for all $q$, the model complexity is controlled by the number of boosting iterations for each distribution parameter. That means that the numbers of boosting iterations are used as the only tuning parameters.
The vector of stopping iterations is determined by resampling methods like cross-validation or bootstrapping using the weights $w_i$ for the observations.
\cite{mayr2012} distinguish between one-dimensional early stopping, that is $m_{\text{stop}}^{(q)} \equiv m_{\text{stop}}$ for $q=1\ldots, Q$, and multi-dimensional early stopping where the stopping iterations $m_{\text{stop}}^{(q)}$ differ for $q=1\ldots, Q$. Multi-dimensional early stopping increases the computational effort, as the optimal stopping iterations are searched on a $Q$-dimensional grid. In the following, we will use multi-dimensional early stopping as it allows for different model complexities for each distribution parameter. Stopping the algorithm early achieves shrinkage of the parameter effects and variable selection, as in each step only the best fitting base-learner is updated. 

For low-dimensional data settings, i.e., ``$N>p$'', and unpenalized estimation, the solution of the boosting algorithm converges to the same solution as maximum likelihood estimation if the number of boosting iterations goes to infinity for all distribution parameters.   
This can be shown using gradient descent arguments \citep{rosset2004,mayr2012}.

\section{Estimation based on penalized maximum likelihood}
\label{sec.penloglike}
We review two estimation approaches that can be used to estimate GAMLSS by maximizing the penalized likelihood.
In contrast to component-wise boosting, where each effect is modeled in a separate base-learner, for penalized likelihood estimation all effects of the linear predictors are estimated together using one likelihood.
We first introduce some notation for this purpose.
The model coefficients $\mtheta_j^{(q)}$, $j=1,\ldots,J^{(q)}$, for the $q$th distribution parameter are concatenated to the vector $\mtheta^{(q)}$, which are concatenated to the vector $\mtheta$ containing the model parameters of all linear predictors.
Analogously, the penalty matrix of all coefficients of the $q$th distribution parameter is $\mP^{(q)}(\mlambda)=\blockdiag( [\mP_j^{(q)}(\mlambda)]_{ j=1,\ldots,J^{(q)}})$, and the penalty matrix for the model coefficients of all linear predictors is $\mP({\mlambda})=\blockdiag( [\mP^{(q)}(\mlambda)]_{q=1,\ldots,Q})$. The generalized inverse of the penalty matrix is denoted by $\mP(\mlambda)^{-}$. We call the vector of all smoothing parameters $\mlambda=(\lambda_1^{(1)}, \ldots, \lambda_{J^{(Q)}}^{(Q)})^\top$. 
For fixed smoothing parameters $\mlambda$, the model coefficients~$\mtheta$ are estimated by maximizing the penalized log-likelihood \cite[see, e.g.,][]{rigby2005,wood2015}
\begin{equation}
\label{eq.lpen}
l_p(\mtheta) = l(\mvartheta, y) - \frac{1}{2} \mtheta^\top \mP(\mlambda)\mtheta,
\end{equation}
where $l(\mvartheta,y)=\sum_{i=1}^N \log f(y_i, \vartheta_i^{(1)}, \ldots, \vartheta_i^{(Q)})$ is the log-likelihood of the data given the distribution parameters and the distribution parameters~$\mvartheta$ depend on the model coefficients~$\mtheta$.

Generally, there are two different approaches to find the optimal smoothing parameters~$\mlambda$.
The first approach is to minimize a model prediction error, for example a generalized Akaike information criterion (GAIC) or a generalized cross-validation criterion (GCV).
The second approach is to use the random effects formulation, where the smoothness penalties can be seen as induced by improper Gaussian priors on the model parameters $\mtheta$, as $\mtheta\sim N(\boldsymbol{0},\  \mP(\mlambda)^{-})$.
The smoothing parameters~$\mlambda$ can then be estimated by maximizing the marginal likelihood with respect to the smoothing parameters. The marginal likelihood is obtained by integrating the model parameters~$\mtheta$ out of the joint density of the data and the model parameters \citep{patterson1971,wood2015},
\begin{equation}
\label{eq.margLik}
\mathcal{V}_r(\mlambda)=\int \exp \left[l(\mvartheta,y) \right] f_{\lambda}(\mtheta) \, \mathrm{d}\mtheta,
\end{equation}
where $f_{\lambda}(\mtheta)$ is the density of the Gaussian prior $N(\boldsymbol{0},\mP(\mlambda)^{-})$. For the normal distribution, maximizing this marginal likelihood is equivalent to maximizing the restricted likelihood (REML). A common approach for maximization is to approximate the integral by Laplace approximation, resulting in the Laplace approximate marginal likelihood (LAML).


\subsection{The gamlss algorithm using backfitting}
\label{sec.rs} 
\cite{rigby2005} first introduced the model class GAMLSS and proposed several variants of a backfitting algorithm for estimation. They provide an implementation for many different response distributions, which at the moment lacks the possibility to specify effects of functional covariates. However, the implementation in \textsf{R}-package \pkg{gamlss} \citep{gamlss2016} allows to specify linear functional effects~\eqref{eq.funCov} as proposed by \cite{wood2011} using the \pkg{gamlss.add}-package \citep{gamlss.add2015} to incorporate smooth terms of the \pkg{mgcv}-package \citep{mgcv2016}. 

In the fitting algorithm, iterative updates of the smoothing parameters~$\mlambda$ and the model coefficients~$\mtheta$ are computed. 
Estimation of the smoothing parameters~$\mlambda$ is done by maximizing the marginal likelihood or minimizing a model prediction error, e.g.~GAIC or GCV, over~$\mlambda$, see Appendix~A of their paper for details. 
For fixed current smoothing parameters $\mlambda$, the model coefficients~$\mtheta$ are estimated using one of two gamlss-algorithms, which are both based on Newton-Raphson or Fisher scoring within a backfitting algorithm. 
Essentially, the algorithms cycle over all distribution parameters $\vartheta^{(q)}$, $q=1,\ldots, Q$, fitting the model coefficients of each distribution parameter in turn by backfitting conditionally on the other currently fitted distribution parameters.
The first gamlss-algorithm is the RS-algorithm which is based on \cite{rigby1996} using first and second derivatives of the penalized log-likelihood with respect to the distribution parameters $\mvartheta$ and is suitable for distributions with information orthogonal parameters, i.e.~the cross-derivatives of the log-likelihood are zero. This is the case, e.g., for the negative binomial, the (inverse) Gaussian and the gamma distribution.
The second gamlss-algorithm is the CG-algorithm which is a generalization of the algorithm introduced by \cite{cole1992} and uses first, second and cross-derivatives of the penalized log-likelihood with respect to the distribution parameters $\mvartheta$. The CG-algorithm is computationally more expensive than the RS-algorithm, see Appendix~B of \cite{rigby2005} for details on both algorithms.

More recently, \cite{rigby2014} proposed a method to estimate the smoothing parameters~$\mlambda$ within the RS- or CG-algorithm using the random effects formulation of penalized smoothing. 
This so called local maximum likelihood estimation is computationally much faster than the previous methods to compute optimal values for~$\mlambda$.


\subsection{Laplace Approximate Marginal Likelihood (LAML) with nested optimization}
\label{sec.mgcv}
\cite{wood2015} developed a general framework for regression with (non-)exponential family distributions, where GAMLSS are contained as a special case. The current implementation supports a Gaussian location scale, a two-stage zero-inflated Poisson model and a multinomial logistic model. Note that in the Gaussian location scale model the inverse standard deviation is modeled.  
Here it is straightforward to use the linear functional effects proposed by \cite{wood2011} for modeling functional covariates as both methods are implemented in \textsf{R}-package \pkg{mgcv} \citep{mgcv2016}. 

The smoothing parameters~$\mlambda$ and the model coefficients~$\mtheta$ are optimized in a nested optimization approach, with an outer Newton optimization to find $\mlambda$ as maximum of the marginal likelihood~\eqref{eq.margLik} and with an inner optimization algorithm to find $\mtheta$ as maximum of the penalized log-likelihood~\eqref{eq.lpen}. 
The integral in the marginal likelihood~\eqref{eq.margLik} is approximated by Laplace approximation resulting in the LAML
\begin{equation*}
\mathcal{V}(\mlambda) = l_p(\hat{\mtheta})
 + \frac{1}{2} \log| \mP(\mlambda) |_{+}
 - \frac{1}{2} \log |\mH | + \frac{M_p}{2} \log(2\pi),
\end{equation*}
where $l_p(\hat{\mtheta})$ is the penalized log-likelihood at the maximizer $\hat{\mtheta}$, $\mH$ is the negative Hessian of the penalized log-likelihood, $| \cdot |_{+}$ denotes a generalized determinant (product of the non-zero eigenvalues) and $M_p$ is the number of zero-eigenvalues of $\mP(\mlambda)$.
For details on the algorithm and stable computations of the necessary components, especially the (generalized) determinant computations see \cite{wood2015}.

For model selection, a corrected AIC is derived by using an adequate approximation of the effective degrees of freedom in the penalized model.
Moreover, it is possible to use the term selection penalties proposed by \cite{marra2011} to do model selection for smooth effects as part of the smoothing parameter estimation.

\section{Comparison of estimation methods}
\label{sec.compare}
In Table~\ref{tab.overview} an overview on the characteristics of the three proposed estimation methods is given, summarizing properties of the gradient boosting algorithm \citep{mayr2012}, cf.~Section~\ref{sec.boosting}, and the two likelihood based approaches, gamlss \citep{rigby2005,rigby2014} and mgcv \citep{wood2015}, cf.~Section~\ref{sec.penloglike}.

\begin{threeparttable}[ht]
\caption{\label{tab.overview}Comparison table for characteristics of the three proposed estimation methods for GAMLSS with functional covariates: component-wise gradient boosting, penalized maximum likelihood based GAMLSS-algorithm using backfitting (gamlss) and using LAML (mgcv).}
\centering 
\begin{tabular}{p{.36\textwidth}| p{.18\textwidth} p{.18\textwidth} p{.18\textwidth}}
Characteristic & boosting & gamlss  & mgcv \\
\hline
response distributions &  many &  many & three$^1$ 
\\
effects of scalar covariates
   & many
   & many
   & many  \\
effects of 
   & linear and 
   & linear$^2$
   & linear$^2$  \\ 
\hspace{1em} functional covariates  & interaction 
   & 
   &   \\ 
types of spline bases
   & P-splines
   & many$^3$
   & many$^3$ \\
in-build variable selection  & yes & no$^4$ & no$^4$ \\
high dimensional data, ``$N<p$'' & yes & no & no \\ 
inference based on & bootstrap & mixed models /  & mixed models / \\
     &   & empirical Bayes  & empirical Bayes \\
computational speed &  &  & \\
\hspace{1em} for large $N$, $p$ & good  & poor & fair \\
\hspace{1em} for small $N$, $p$ & fair  & fair & good \\
\end{tabular}
\begin{tablenotes}
\footnotesize
\item $^1$ Gaussian location scale, two-stage zero-inflated Poisson, multinomial logistic model
\item $^2$ built in implementation only for functional covariates with observation grids that imply integration weights~1, e.g.~$(s_1, \ldots, s_R)^\top = (1, \ldots, R)^\top$; use $\tilde{x}(s) = \Delta(s) x(s)$ to estimate coefficient functions for covariates observed on differently or unevenly spaced grids 
\item $^3$ e.g., P-splines, thin plate splines and adaptive smoothers
\item $^4$ variable selection e.g.~based on information criteria possible
\end{tablenotes}
\end{threeparttable}
\vspace*{1em}

For the response distribution the gamlss and the boosting approach provide the same flexibility, whereas the mgcv approach currently only has the possibility to specify three different distributions. 
Regarding the modeling of functional covariates, the boosting approach is the most flexible, as it allows to specify interaction effects between scalar and functional covariates.
All three methods allow for a large variety of covariate effects of scalar covariates, including for example smooth, spatial and interaction terms.
Using boosting, it is possible to estimate models in high dimensional data settings with many covariates, where maximum likelihood methods are infeasible. Using maximum likelihood based methods, inference is a byproduct of the mixed model framework, providing confidence intervals and p-values. In the boosting context, inference can be based on bootstrapping or permutation tests \citep{mayr2015}. 
Comparing the computational speed, gamlss and mgcv are considerably faster than boosting for small data settings, as boosting requires resampling to determine the optimal number of boosting iterations. For many observations and especially for many covariate effects,  boosting scales better than the likelihood-based methods, as it fits each base-learner separately.

\section{Model choice and diagnostics}
\label{sec.modelchoice}
In practical applications of GAMLSS one is faced with several decisions on model selection concerning not only the choice of the response distribution but also the choice of the relevant covariates for each distribution parameter. We will shortly discuss normalized quantile residuals \citep{dunn1996} and global deviance (GD).
These tools are particularly suited to different model selection tasks. Quantile residuals can be used as a graphical tool to check the data fit under a certain response distribution. The GD can be used to measure how closely the model fits the data.
\cite{rigby2005} and \cite{wood2015} both discuss a generalized Akaike information criterion (GAIC) which penalizes the effective degrees of freedom of the model and can be used to compare non-nested and semi-parametric models. For the boosting approach no reliable estimates for the degrees of freedom are available. The GAIC and other information criteria are only comparable for models that are estimated by the same estimation method and thus will not be used in the following. 
\\\\
\underline{Quantile residuals.}
Quantile residuals can be used to check the adequacy of the model and especially of the assumed response distribution. For continuous responses, quantile residuals are defined as $\hat{r}_i=\Phi^{-1}(v_i)$, where $\Phi^{-1}$ is the inverse distribution function of the standard normal distribution, and $v_i=F(y_i|\hat{\mvartheta}_i)$. Here $F$ is the distribution function of the assumed response distribution and $\hat{\mvartheta}_i$ are the predicted distribution parameters $(\hat{\vartheta}^{(1)}_i, \ldots, \hat{\vartheta}^{(Q)}_i)^\top$ for the $i$th observation.
For discrete integer valued responses, normalized randomized quantile residuals can be used \citep{dunn1996}.
In the special case of the normal distribution, the computation of the quantile residuals can be simplified to $\hat{r}_i=(y_i - \hat{\mu_i})/\hat{\sigma_i}$.
If the distribution function $F(\cdot|\hat{\mvartheta}_i)$ that is predicted by the model is close to the true distribution, the quantile residuals follow approximately a standard normal distribution. This can be checked in quantile-quantile plots (QQ-plots).
\\\\
\underline{Global deviance.}
The fitted global deviance (GD) is defined as GD$=-2l(\hat{\mvartheta},y)$, with $l(\hat{\mvartheta},y)=\sum_i l(\hat{\mvartheta}_i,y_i)$, and measures how closely the model fits the data. 
\cite{mayr2012} propose to use the empirical risk for model evaluation which for GAMLSS is the negative log-likelihood, $-l(\hat{\mvartheta},y)$, and therefore is equivalent to the GD. 
To avoid over-fitting, we compute the GD on test-data that were not used for the model fit. Thus, the GD can be used to compare models fitted under different distributional assumptions, by different algorithms and using different linear predictors.


\section{Financial returns of Commerzbank stock}
\label{sec.stocks}
In this section, we apply our model to the motivating time series of daily stock returns for Commerzbank shares as recorded by the XETRA electronic trading system of the German stock exchange.
The log-returns are defined as $y_i=\log(p_{i1}/p_{i0}) \approx (p_{i1} - p_{i0}) / p_{i0} $, where $p_{i0}$ is the price at opening and $p_{i1}$ is the price at closing of day~$i$. We use data from November 2008 to December 2010 ($N=527$).
In addition to the price of the shares, our data also provide information about supply and demand, i.e., the liquidity of the stock, over a trading day. The role of liquidity within the price formation is a question of major interest in finance and economics \citep{amihud2002,amihud2013}.

Traditionally, stock returns have been modeled by pure autoregressive specifications which already capture their major stylized facts: While being weakly serially correlated, their conditional variance is strongly serially dependent. Moreover, the unconditional distribution is fat-tailed. The autoregressive conditional heteroskedasticity (ARCH) model of \cite{engle1982} captures all these features. It is defined as
\begin{align}
y_i &= \sigma_i \varepsilon_i,
\text{ with } \varepsilon_i \stackrel{iid}{\sim} N(0,\ 1),\nonumber \\
\sigma_i^2 &= \beta_0 + \sum_{j=1}^{p} \beta_j y_{i-j}^2,
\end{align}
where $\beta_0 > 0$, $\beta_j \geq 0$, $j = 1, \ldots, p$, and $i$ is the time index. Hence, the distribution of $y_i$ conditional on the $p$ lagged returns is given by a normal distribution with mean zero and variance $\beta_0 + \sum_{j=1}^{p} \beta_j y_{i-j}^2$, and the model is called (linear) ARCH($p$).
In the following, we avoid the aforementioned nonnegativity-constraints on $\beta_j$ by using a log-link. 
Additionally allowing for $p_1$ autoregressive (AR) effects in the conditional mean, we arrive at 
\begin{align}
y_i &\sim N\left(\alpha_0 + \sum_{j=1}^{p_1} \alpha_j y_{i-j},\
\exp \left[ \beta_0 + \sum_{j=1}^{p_2} \beta_j y_{i-j}^2 \right] \right),
\label{eq.ARCH}
\end{align}
which can be viewed as a generalized linear model for location and scale, with Gaussian response distribution, identity link for the expectation and log-link for the variance.

\begin{figure}[ht]
\centering
\makebox{
  \includegraphics[width=0.3\textwidth, page=1]{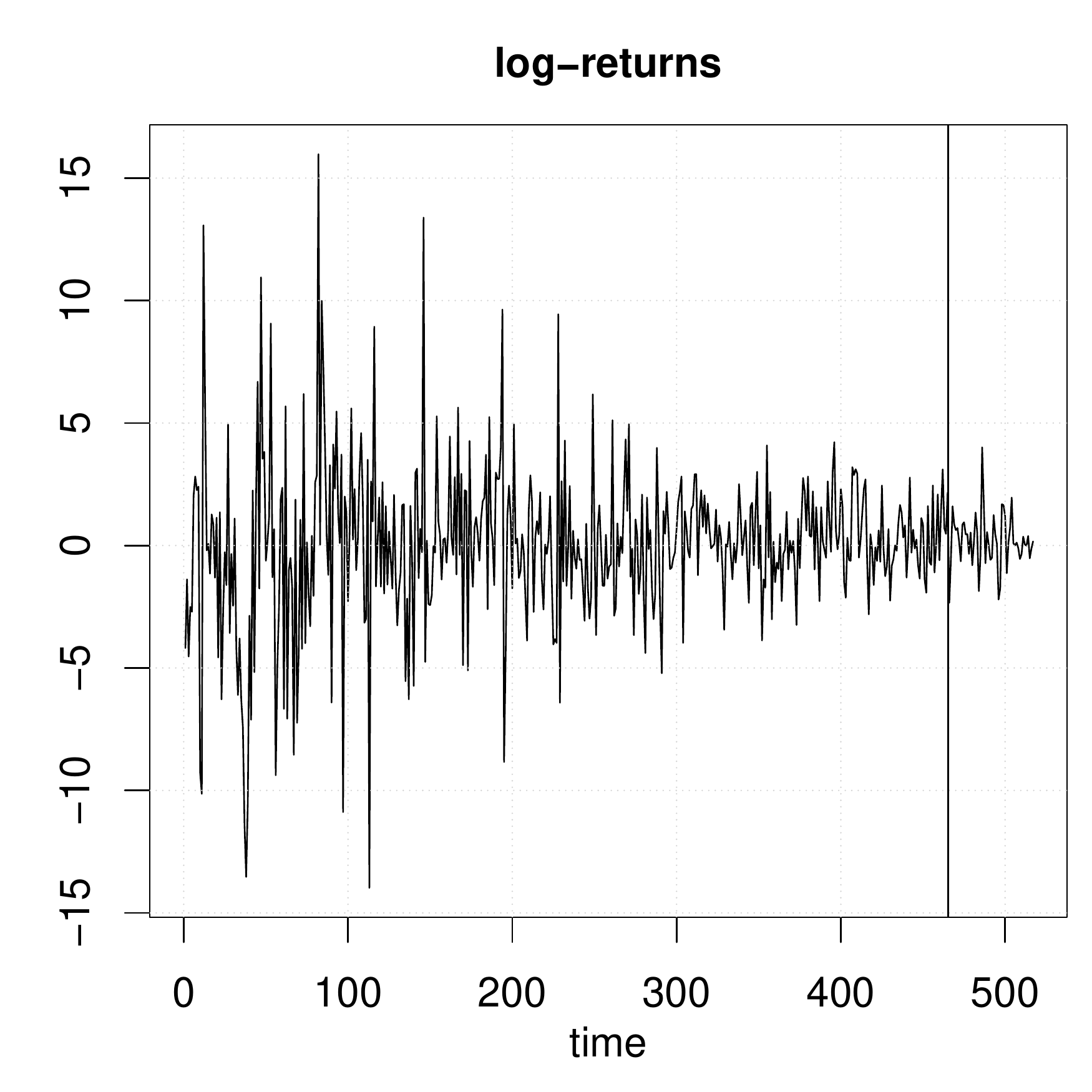}
  \includegraphics[width=0.3\textwidth, page=2]{logReturns.pdf}
  \includegraphics[width=0.3\textwidth, page=3]{logReturns.pdf}
  }
\caption{\label{fig.stocks}Descriptive plots of the data in the application: The open-to-close returns on the mid-quote for the stocks of Commerzbank from November 2008 to December 2010 (left), the averaged cumulative liquidity curves for the offered (middle) and requested (right) number of shares. The vertical line in the left plot indicates the split into training and test data. 
}
\end{figure}

At each point in time during a given trading day, the XETRA system records all outstanding limit orders, i.e., offers and requests to sell or buy a certain number of shares at a specified price which are not immediately executed against a suitable order of the opposite market side. By construction, prices of these offers (requests) are above (below) the current price, the mid-price, which is defined as the mean of the best (i.e., lowest) offer and the best (i.e., highest) request. 
Our data set contains for each trading day and both market sides the mean number of shares or volumes at a distance of $0,\ldots,200$ Cents to the current market price. The mean over the trading day is computed based on snapshots of the order book taken every 5 minutes during the trading hours (9am to 5:30pm).
From this information, functional measures of liquidity can be constructed \citep{haerdle2012,fuest2015}.
Cumulating the volumes along the price axis -- with increasing (decreasing) price on the supply (demand) side --, one obtains non-decreasing curves: the cumulative volume in the market as functions of the relative price. The relative price is standardized to $s \in [0,1]$. See Figure~\ref{fig.stocks} for descriptive plots of the data.

The cumulative volume curves can be viewed as measures of liquidity: Liquidity is high if the curves are steep, and it is low if they are flat. However, the curves are nonlinear in general.

\subsection{Model choice}
The functional regression approach put forth in this paper enables us to estimate the impact of functional liquidity on the stock returns' conditional location and scale parameters in a very general way.
We consider the normal and the $t$-distribution as possible response distributions in the GAMLSS and use the methods highlighted in Section~\ref{sec.modelchoice} to select models. For the model fit, we use the first 90\% of the time series as training data, and keep the remaining 10\% as test data to evaluate the model fits computing the GD out-of-bag.

Assuming normally distributed response, we specify the following model for the returns $y_i$ depending on lagged response variables and the two functional liquidities $x_{i,\text{ask}}$ and $x_{i,\text{bid}}$, with $i=1, \ldots, N$: 
\begin{equation}
\begin{split}
y_i| y_{i-1}, \ldots, y_{i-\max(p_1,p_2)}, x_{i,\text{ask}}, x_{i,\text{bid}}  \sim N(\mu_i,\ \sigma_i^2),  \\
\mu_i = h^{(\mu)}(x_i) = \alpha_0 + \sum_{j \in \{\text{ask}, \text{bid}\}} \int x_{ij}(s)\alpha_j(s) \, \mathrm{d}s + \sum_{j=1}^{p_1} \alpha_j y_{i-j}, \\
\log \sigma_i = h^{(\sigma)}(x_i) =\beta_0 + \sum_{j \in \{\text{ask}, \text{bid}\}} \int  x_{ij}(s) \beta_j(s) \, \mathrm{d}s  + \sum_{j=1}^{p_2} \beta_j \log y_{i-j}^2.
\end{split}
\label{eq.stockModel}
\end{equation}
We use $p_1=p_2=10$ lagged variables for the expectation and the standard deviation leaving us with $N=527 - 10 = 517$ observations.
To obtain coefficients for the variance instead of the standard deviation, the corresponding model equation is multiplied with two, as $\log \sigma_i^2 = 2\log \sigma_i$. 
In \textsf{R} package \pkg{mgcv} the parameterization of the model is slightly different from~ \eqref{eq.stockModel}, as the scale parameter is the inverse standard deviation $\tau_i = 1/ \sigma_i$ which is modeled using $\log(1/\tau_i - \epsilon)$ as link function, where $\epsilon$ is a small positive constant is used to prevent that the standard deviation tends to zero. We use $\epsilon = 0.01$.  Reformulating the mgcv parameterization yields $\log(\sigma_i - \epsilon) = h^{(\sigma)}(x_i)$ and thus a very similar interpretation for all coefficients.    

Model~\eqref{eq.stockModel} nests the purely autoregressive location scale specification in equation~\eqref{eq.ARCH} that is common in the econometric literature. 
Results for the autoregressive parameters suggest that returns exhibit only a very weak serial dependence, which is in line with the findings typically reported in the literature.
For the variance equation, the overwhelming majority of empirical studies uses the GARCH (generalized ARCH) model of \cite{bollerslev1986} which includes just one lagged squared return, but additionally $\sigma_{i-1}^2$ as latent covariate and is called GARCH(1,1). As $\sigma_{i-1}^2$ cannot be observed, this approach is not nested in our model class. However, it can be shown that GARCH(1,1) can be represented as an ARCH($\infty$) process with a certain decay of the autoregressive parameters. 
As the boosting algorithm employed in our approach implicitly selects relevant covariates, we do not have to impose such a restrictive structural assumption. Instead, we allow for a generous number of lags (10 lags), finding that only the first few (around 5 lags, cf.~Figure~\ref{fig.coef}) are non-zero. 

As a more heavy-tailed alternative to the normal distribution we specify a three parameter Student's $t$-distribution, $Y\sim t(\mu,\ \sigma,\ \df)$, with location parameter~$\mu$, scale parameter~$\sigma$ and degrees of freedom $\df$ \citep{lange1989}. This implies $\EV(Y)=\mu$, for $\df>1$, and $\text{sd}(Y)=\sigma\sqrt{\text{df}/(\text{df} - 2)}$, for $\df>2$. We specify the linear predictors for the parameters $\mu$ and $\sigma$ as in model~\eqref{eq.stockModel} and model $\df$ as constant. 

The scalar effects of the lagged responses are estimated as linear effects without penalty. 
The effects of the functional covariates $x_{ij}(s)$ are specified as in~\eqref{eq.funCov} using 20 cubic B-splines with first order difference penalty matrices, resulting in P-splines \citep{eilers1996} for $\alpha_j(s)$, $\beta_j(s)$. 
Alternatively we use the FPC basis functions to represent the functional covariates and  functional coefficients, yielding a regression onto the scores as in equation~\eqref{eq.fpcaeffect}. We choose the first $3$ FPCs as those explain 99\% of the variability in the bid- and the ask-curves.  

\begin{figure}[ht]
\centering
\makebox{
\raisebox{0.2\textwidth}{sc\phantom{+fun }}
\includegraphics[width=0.34\textwidth, page=1]{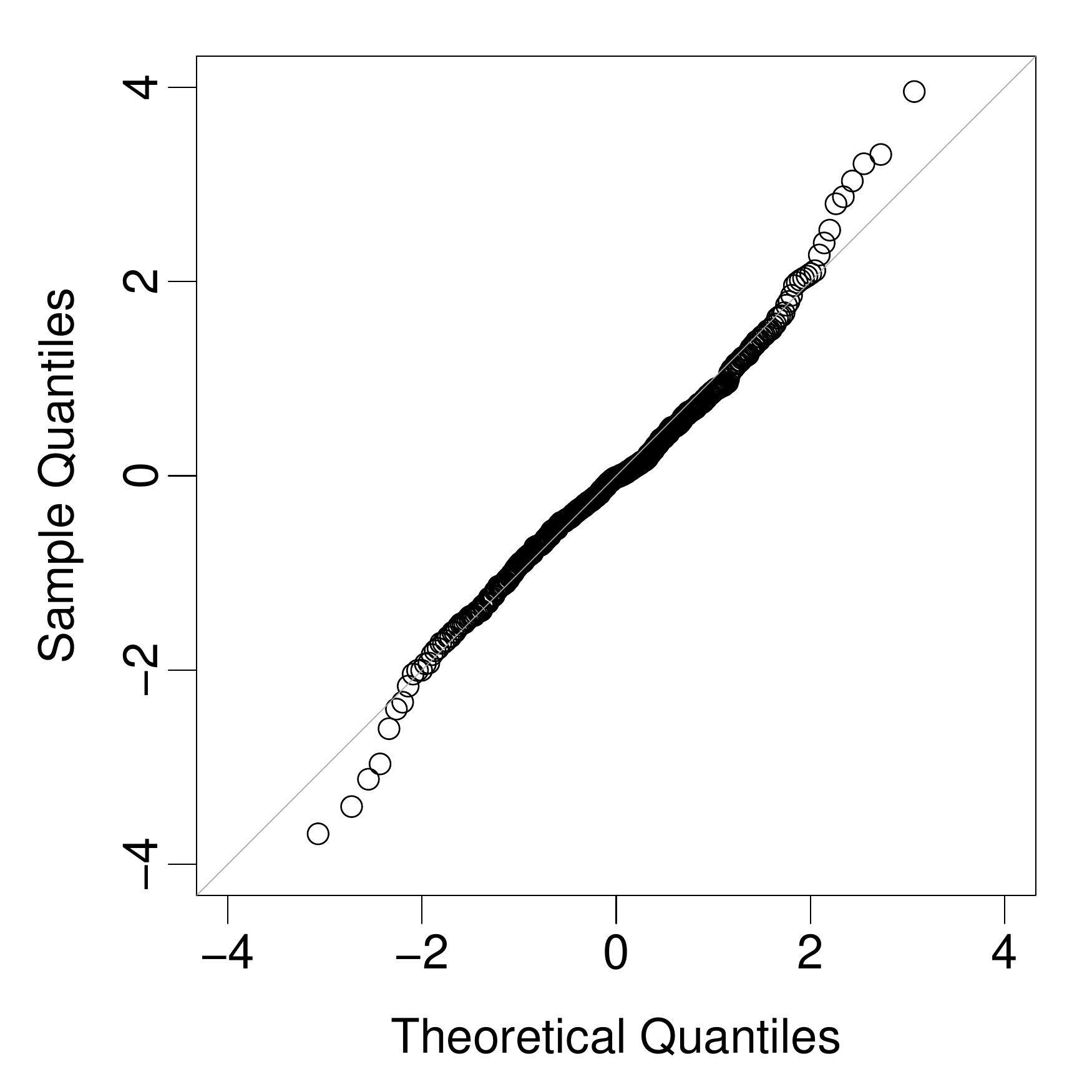}
\includegraphics[width=0.34\textwidth, page=1]{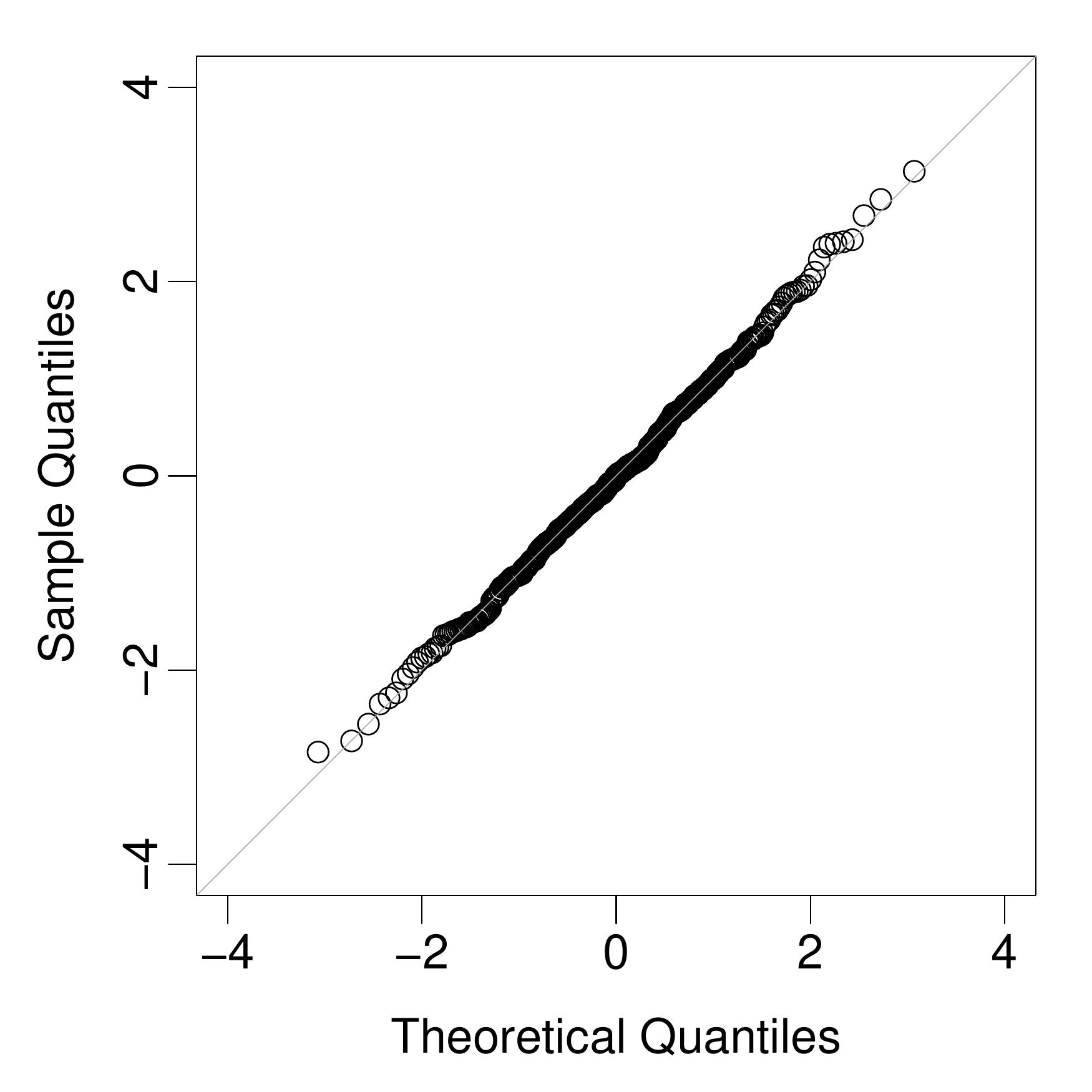}
 }
\makebox{
 \raisebox{0.2\textwidth}{sc+fun  }
\includegraphics[width=0.34\textwidth, page=1]{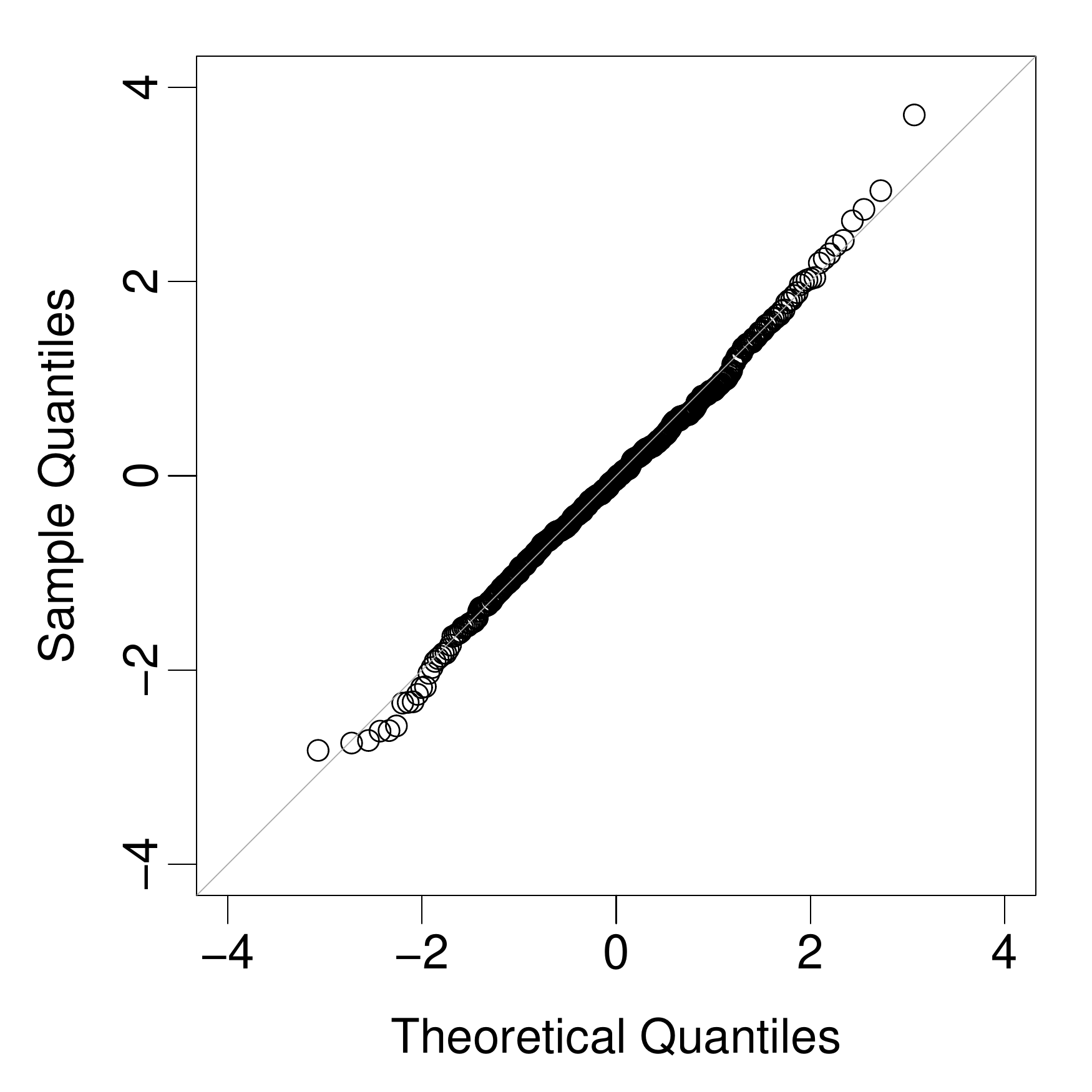}
\includegraphics[width=0.34\textwidth, page=1]{student_gamlss_sc_90_inbag_qq_acf.pdf}
 }
\caption{\label{fig.qqplot}Model choice for the application: QQ-plots of the quantile residuals in the GAMLSS with scalar variables (sc, top row) or with scalar and functional variables (sc+fun, bottom row) fitted by P-splines; 
assuming normally distributed response (left column) or Students $t$-distribution for the response (right column) for models estimated by the gamlss algorithm. The diagonal is marked by a grey line.}
\end{figure}

For the estimation by boosting, we use 100 fold block-wise bootstrapping \citep{carlstein1986} with block length 20 to find the optimal stopping iterations. We search on a two dimensional grid, allowing different numbers of boosting iterations for the distribution parameters. As the third distribution parameter of the $t$-distribution, df, is modeled as constant, we only use a two-dimensional grid, setting the number of iterations for df to the maximum value of the grid.
For the normal distribution the step-lengths are fixed at $(0.1, 0.01)^\top$, for the $t$-distribution at $(0.1, 0.01, 0.1)^\top$, as the boosting algorithm was found to run more stably with smaller step-lengths for variance parameters. 
For the likelihood-based estimation methods we use the same design and penalty matrices, but the smoothing parameters $\mlambda$ are estimated by a REML or LAML criterion. 
\\\\
\underline{Quantile residuals.}
As the QQ-plots look similar for all three estimation methods we here only show them for one method. In Figure~\ref{fig.qqplot} the QQ-plots of the quantile residuals for the models assuming normally and $t$-distributed response fitted by gamlss are given. As covariates the lagged response values or the lagged response values and the liquidity curves are used.
The QQ-plots indicate that all models fit the data reasonably well for residuals in $[-2,2]$. More extreme residuals are better captured by the $t$- than by the normal distribution. The QQ-plot for the normal model with lagged response values as covariates is s-shaped indicating that the sample quantiles are heavier tailed than the true quantiles. For the model using in addition the liquidity curves the normal distribution seems to be adequate. Thus, it seems that the functional covariates can better explain the extreme values such that the heavy tails of the $t$-distribution are not necessary.

As we analyze time-series data we check the squared residuals for serial correlation by looking at the estimated autocorrelation function (ACF), see Figure~\ref{fig.acf} in Appendix~\ref{sec.appendix_stocks}.  
For lags greater one the autocorrelation is close to zero.
\\\\
\underline{Comparison by global deviance.}
We compare the GD, standardized by the number of observations in the test data,~$N_{\text{test}}$, for models fitted with the three estimation methods and assuming normally or $t$-distributed response, see Table~\ref{tab.pred}.
To check for the benefit of using the functional liquidity curves as covariates, we compute models using only the lagged response values for comparison.
We compute the GD on the last 10\% of the time-series using two different procedures. We do one-step predictions, refitting the model on the data up to the time-point $i-1$ and predict the distribution parameters $\hat{\mvartheta}_i$ using this model. Alternatively we do the predictions using the model on the first 90\% of the data.
\begin{table}[ht]
\caption{\label{tab.pred}Results for the application: Goodness of the model fit measured by the general deviance (GD) of the models assuming normal or $t$-distribution; estimated by boosting, gamlss and mgcv; using the lagged response values as covariates (sc), or the lagged response values and the functional liquidity curves (sc+fun). The functional terms are estimated using P-splines (P) or an FPCA-basis (FPCA).
The GD is computed on refitted models up to the day that should be predicted (one-step) or on the model using the first 90\% of the data. }
\centering
\begin{tabular}{ll|ccc|ccc} 
 & & \multicolumn{3}{c}{GD/$N_{\text{test}}$ (one-step)} & \multicolumn{3}{|c}{GD/$N_{\text{test}}$ }\\
 & & sc & \multicolumn{2}{c|}{sc+fun} & sc  & \multicolumn{2}{c}{sc+fun} \\
 distribution & estimation  & - & P &  FPCA & - & P &  FPCA \\
  \hline
 normal & boosting     & 3.24 & 3.16   & 3.10 & 3.30 & 3.27  & 3.19 \\
 normal & gamlss       & 3.20 & 2.98   & 3.09 & 3.26 & 3.15  & 3.19 \\
 normal &  mgcv        & 3.20 & 2.97   & 3.09 & 3.26 & 3.14  & 3.19 \\
 \hline
  $t$ & boosting       & 3.46 & 3.39   & 3.35 & 3.06 & 3.11  & 3.08 \\
  $t$ & gamlss         & 3.09 & 2.95   & 3.06 & 3.15 & 3.11  & 3.16 \\
\end{tabular}
\end{table}
The GD of the models fitted by mgcv and gamlss is very similar and mostly smaller than that for the models fitted by boosting. 
For the GD computed on the models for the first 90\% of the data, the models assuming $t$-distribution outperform those with normal distribution and there is no or not much additional advantage of the models using the functional covariates in addition to the scalar lagged covariate effects. 
Using the functional liquidity terms in addition to the lagged scalar covariates improves the one-step GD and to a smaller extent the GD, especially for models assuming normally distributed response. The models using P-splines for the functional effects have smaller GD than those using FPCA when fitted with mgcv or gamlss.


\subsection{Results}
\label{sec.stocks_results}
\begin{figure}[ht]
\centering
\mbox{
 \includegraphics[width=0.3\textwidth, page=1]{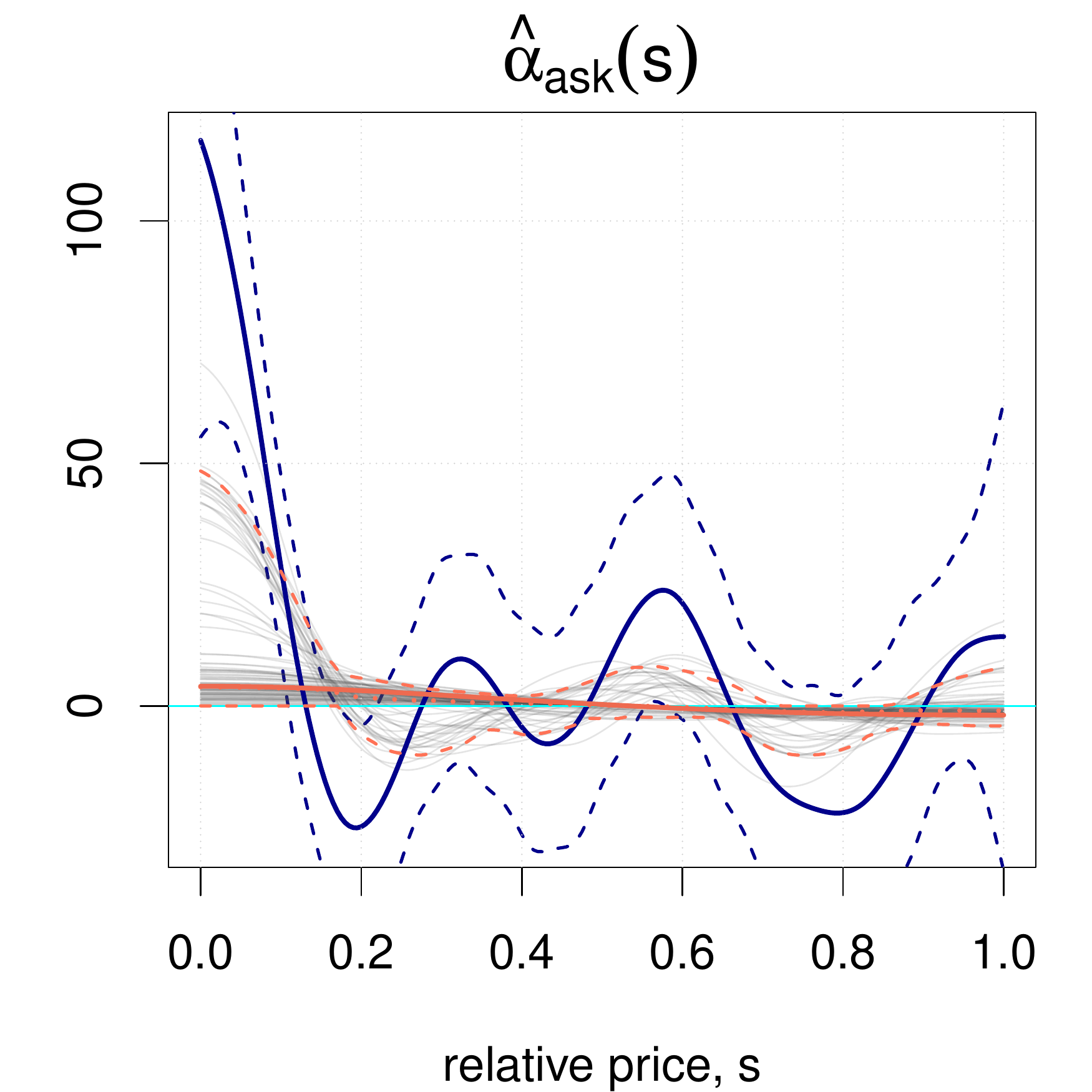}
 \includegraphics[width=0.3\textwidth, page=2]{Gaussian_boost_mgcv_90_intercept_coefBS.pdf}
 \includegraphics[width=0.3\textwidth, page=3]{Gaussian_boost_mgcv_90_intercept_coefBS.pdf}
 }
\mbox{
 \includegraphics[width=0.3\textwidth, page=4]{Gaussian_boost_mgcv_90_intercept_coefBS.pdf}
 \includegraphics[width=0.3\textwidth, page=5]{Gaussian_boost_mgcv_90_intercept_coefBS.pdf}
 \includegraphics[width=0.3\textwidth, page=6]{Gaussian_boost_mgcv_90_intercept_coefBS.pdf}
 }
\caption{\label{fig.coef}Results for the application, model with normal distribution: Estimated coefficients for $\mu_i$ (top panel) and $\sigma_i$ (bottom panel) in the GAMLSS with the two liquidities as functional covariates and $p_1=p_2=10$ lag variables.
For the intercept of the standard deviation we plot $\hat{\beta}_0 - 1$ to better fit the intercept into the range of the lag effects.
The boosting estimates on the 100 block-bootstrap samples are plotted as partly transparent lines or circles and the point-wise 2.5, 50, and 97.5\% quantiles as dashed orange lines. The boosting estimates are plotted as solid orange line.
The estimates of mgcv with point-wise 95\% confidence bands are plotted in dark blue. The zero-line is marked with a light-blue line.}
\end{figure}

The fitted coefficients for the models assuming normal or $t$-distribution are quite similar. We show the estimated coefficients for the normal location scale model with functional liquidity effects estimated using P-splines, and refer to Appendix~\ref{sec.appendix_stocks} for further results. For the corresponding model assuming $t$-distributed response, the predicted df are $\exp(\hat{\gamma}_0) \approx 8.2$, with 95\% confidence interval $[5.0, 13.3]$ for gamlss and $\exp(\hat{\gamma}_0) \approx 3.8$ with 95\% bootstrap confidence interval $[3.3, 7.7]$ for boosting. 
The estimated coefficients for the normal location scale model \eqref{eq.stockModel} fitted by boosting and by mgcv can be seen in Figure~\ref{fig.coef}. As the estimates by gamlss and mgcv are very similar we only show the results for one of the likelihood-based methods. 
The parameter estimates of the autoregressive parts in both the expectation and standard deviation equation imply stationary dynamics, that means the time-series induced by the lagged scalar effects are stationary.
For boosting, the estimated coefficients on the 100 bootstrap-samples are depicted together with point-wise 2.5, 50, and 97.5\% quantiles. For the mgcv-approach the estimated coefficient functions are given with point-wise 95\% confidence intervals.
The estimated coefficients for the standard deviation are quite similar. 
Regarding the effects on the expectation, the size and smoothness of the estimated coefficients obtained by mgcv and by boosting differs considerably, as the coefficients obtained by boosting are  shrunken towards zero. 

In a simulation study, see Section~\ref{sec.simGAMLSS} and Appendix~\ref{sec.sim_appendix}, we observe that depending on how much information the functional covariates contain, the functional coefficients can be estimated with more or less accuracy. The functional covariates in this application contain relatively little information as in an FPCA more than 99\% of the variance can be explained by the first three FPCs and the explained variance per principal component is strongly decreasing. 

When using only a small number of basis functions in the specification of the functional effects in mgcv (e.g., $K_j=10$ instead of 20) and using the shrinkage penalty of \cite{marra2011}, mgcv yields similar estimates like boosting (with $K_j=10$ or 20), see Figure~\ref{fig.coef_shrink10} in Appendix~\ref{sec.appendix_stocks}. 
When the number of boosting iterations is increased, the boosting-estimates become similar to those obtained by mgcv (not shown). Thus, the different results are mainly due to the different selection and estimation of hyper-parameters that imply different choices for the effective degrees of freedom for the smooth effects. 

The effect size depends on the underlying modeling assumptions and we will only interpret the direction of the effects (positive or negative), as those are estimated stably. These directions of the effects remain, even when fitting the models using FPCA-basis functions, although the shape of the functional effects changes, as the effects are assumed to lie in the space spanned by the FPCA-basis functions. This implies for this application that all functional effects start almost in zero (cf.~Figure~\ref{fig.coef_FPCA}).

Looking at the estimated functional coefficients in Figure~\ref{fig.coef}, the absolute values of the estimated coefficient functions are generally higher for small $s$, which is sensible, as the bid and ask curves for small $s$ describe the liquidity close to the mid-price. For the effects on the expectation, the estimates of $\hat\alpha_{\text{ask}}(s)$ are positive and $\hat\alpha_{\text{bid}}(s)$ are negative near the mid-price. Thus, higher liquidity of the ask side and lower liquidity of the bid side tend to be associated with an increase of the expected returns. The lagged response values seem to have no influence on the expectation, as $\hat\alpha_j$ is virtually always zero for the boosting estimation, and close to zero for the mgcv-estimation with confidence bands containing zero.
Looking at the model for the standard deviation, the estimated coefficient functions for ask, $\hat\beta_{\text{ask}}(s)$, are mostly negative. This means that higher liquidity leads to lower variances, and lower liquidity leads to higher variances. The estimated coefficients for the bid curves, $\hat\beta_{\text{bid}}(s)$, are quite close to zero.
The lagged squared response values seem to have an influence for close time-points, as many $\hat\beta_j$ are greater zero for the first five lags.


\section{Simulation studies}
\label{sec.simGAMLSS}
In the following we present a simulation study using the observed functional covariates of the application and coefficient functions resembling the estimated ones. Then we comment on the results of a more general simulation study comparing the estimation methods systematically in different settings.

\subsection{Simulation study for the application}
\begin{figure}[ht]\centering
\mbox{
 \includegraphics[width=0.3\textwidth, page=1]{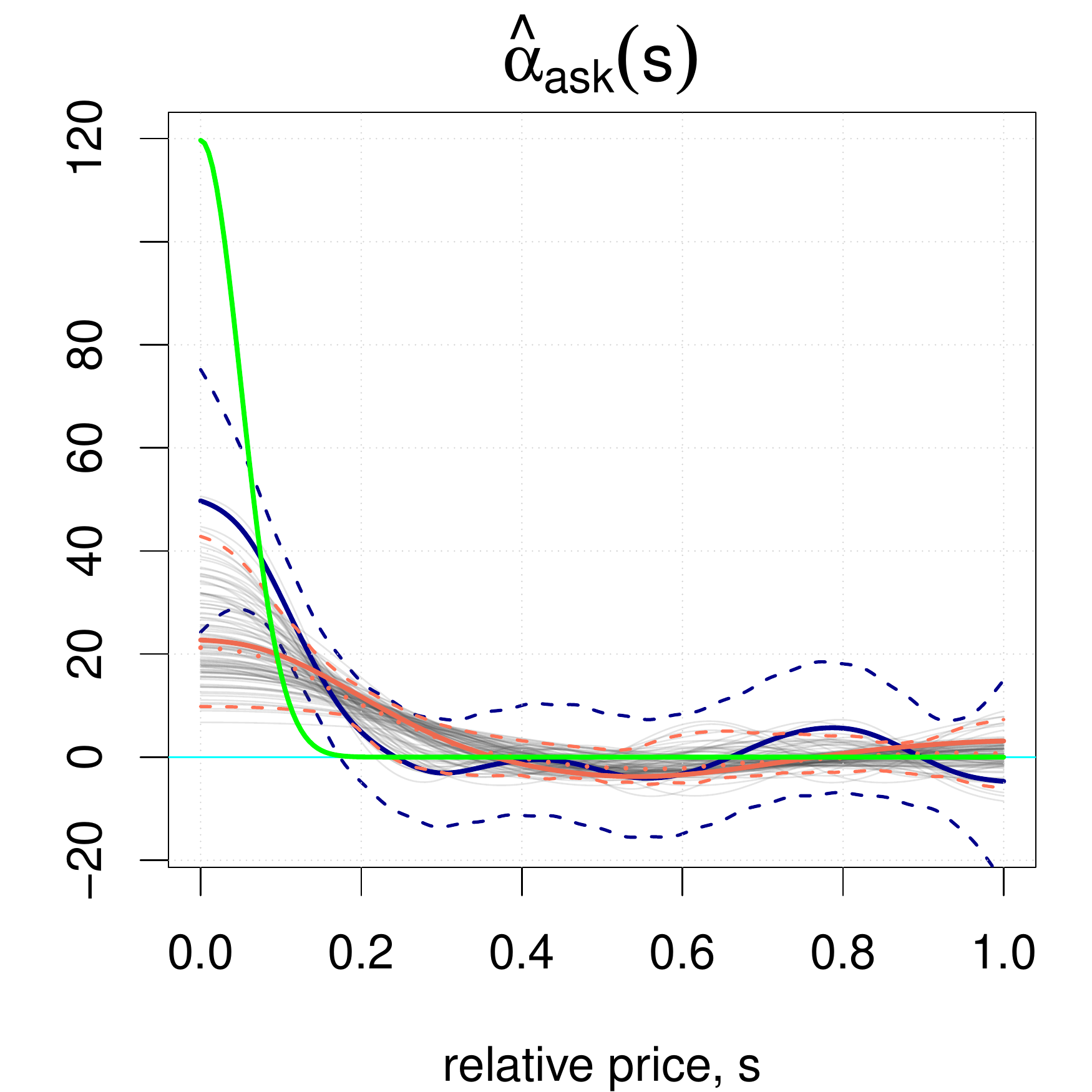}
 \includegraphics[width=0.3\textwidth, page=2]{Gaussian_sim1_boost_mgcv_90_intercept_coefBS.pdf}
 \includegraphics[width=0.3\textwidth, page=3]{Gaussian_sim1_boost_mgcv_90_intercept_coefBS.pdf}
 }
 \mbox{
 \includegraphics[width=0.3\textwidth, page=4]{Gaussian_sim1_boost_mgcv_90_intercept_coefBS.pdf}
 \includegraphics[width=0.3\textwidth, page=5]{Gaussian_sim1_boost_mgcv_90_intercept_coefBS.pdf}
 \includegraphics[width=0.3\textwidth, page=6]{Gaussian_sim1_boost_mgcv_90_intercept_coefBS.pdf}
 }
\caption{Results of the simulation study: Estimated coefficients for $\mu_i$ (top panel) and $\sigma_i$ (bottom panel) in the GAMLSS with the two liquidities as functional covariates and $p_1=p_2=10$ lag variables for simulated response observations. The true underlying coefficients are plotted as green lines.
For the intercept of the standard deviation we plot $\hat{\beta}_0 - 1$ to better fit the intercept into the range of the lag effects.
The boosting estimates on the 100 block-bootstrap samples are plotted as partly transparent lines or circles and the point-wise 2.5, 50, and 97.5\% quantiles as dashed orange lines. The boosting estimates are plotted as solid orange line.
The estimates of mgcv with point-wise 95\% confidence bands are plotted in dark blue. The zero-line is marked with a light-blue line.}
\label{fig.simStock}
\end{figure}

To check the obtained results in the application we conduct a small simulation study using the functional covariates from the real data set to simulate response values using coefficients that are similar to the estimated ones. Then we fit the normal location scale model in the same way as described above and compare the estimated coefficients with the true coefficients that were used for simulating the response, see Figure~\ref{fig.simStock}. The results of mgcv and gamlss are again very similar.
Generally the functional coefficients capture the form of the true coefficients, but they are shrunken towards zero for both estimation methods, with boosting shrinking more strongly. Especially the very high coefficient function for small relative price for the bid liquidity is underestimated, as the functional observations contain only very little information for small $s$ due to all curves starting almost in zero. 
This illustrates that in this particular case of little information content in the functional covariates, while the model is still (just) identifiable, the smoothness assumption encoded in the penalties and the early stopping, which leads to a shrinkage effect, still strongly influences the estimates. The strength of this effect differs between the two estimation approaches, but is in the same direction. For our application, this implies that the true effects might be underestimated, but that the sign of the effects should be interpretable as we assumed. 

\subsection{General simulation study}
\label{sec.genSim}
The aim of the simulation study is to check and compare the model fits of the different implementations systematically. \cite{mayr2012} conduct a simulation study for boosting GAMLSS with scalar covariates. In \cite{wood2015} the performance of mgcv and gamlss is compared for some settings  with scalar covariates. We thus focus on the estimation of effects of functional covariates. For the generation of functional covariates we use different processes resulting in functional covariates containing different amounts of information. Comparing the estimates of the functional effects over those settings shows that the data generating process of the functional covariates strongly influences the estimation accuracy of the functional effects. This has already been discussed in \cite{scheipl2016} for mean models. On the other hand we vary the complexity of the functional coefficients from zero-coefficients over almost linear to u-shaped coefficient functions and functions with a steep bend. With growing complexity of the coefficient functions the estimates tend to get worse. Generally one can observe in the normal location scale model that the coefficient estimates for the expectation are better than those for the standard deviation. 

In the simulation study the most difficult data setting, i.e.\ the one with the least information in the covariates, best reflects the data situation in the application on stock returns. In this case the coefficient estimates often underestimate the absolute value of the coefficients. 
For less pathological data situations the estimates are usually close to the true functions and all three estimation approaches yield very similar estimates and predictions. 
See Appendix~\ref{sec.sim_appendix} for details on the data generation and the results of this simulation study.

\section{Discussion}
\label{sec.discussion}
In this paper we discuss the extension of scalar-on-function regression to GAMLSS. The flexibility of the  approach allows to model many different response distributions with parameter-specific linear predictors, containing effects of functional and scalar covariates. The two proposed estimation methods based on boosting and on penalized likelihood are beneficial in different data settings. The component-wise gradient boosting algorithm can be used in high-dimensional data settings and for large data sets. The likelihood-based methods provide inference based on mixed models and can be computed faster for small data sets.

We believe that the combination of scalar-on-function regression and GAMLSS is an important extension to functional regression models with many possible applications in different fields. In particular, scalar-on-function models for zero-inflated or over-dispersed count data as well as bounded continuous response distributions can be fitted within the GAMLSS framework. In contrast to quantile regression models with functional covariates \cite[e.g.,][]{ferraty2005,cardot2005,chen2012} GAMLSS provide coherent interpretable models for all distribution parameters, prevent quantile crossing and allow for simultaneous inference at the price of assuming a particular response distribution.

In future research, we will consider GAMLSS for functional responses with scalar and/or functional covariates, with the aim of extending the flexible functional additive mixed model framework of \cite{scheipl2015} and \cite{brockhaus2015}, estimated by penalized likelihood and boosting, respectively, to simultaneous models for several response distribution parameters.


\subsection*{Acknowledgments}
The authors thank Deutsche B\"orse AG for providing the \mbox{XETRA} historical limit order book data as part of a cooperation with the Chair of Financial Econometrics, LMU Munich, Germany. The work of Sarah Brockhaus and Sonja Greven was supported by the German Research Foundation through Emmy Noether grant GR 3793/1-1.


\clearpage

\begin{appendix}


\section{Details on the implementation of the estimation methods}

\subsection{Used \textsf{R} packages}
\label{sec.rpackages}
All computations are performed using \textsf{R} software for statistical computing \citep{R2015}. Boosting for GAMLSS, Section~\ref{sec.boosting},  
is implemented in the \textsf{R}~package \pkg{gamboostLSS} \citep{gamboostLSS2015} and base-learners for functional covariates are available in the \pkg{FDboost}-package \citep{FDboost2016}.
The \pkg{gamboostLSS}-package provides the possibility to use all distribution families implemented in the \pkg{gamlss}-package \citep{gamlss2016}.

Estimation of GAMLSS by maximizing the penalized log-likelihood using backfitting, Section~\ref{sec.rs}, is implemented in the \textsf{R}~package \pkg{gamlss} \citep{gamlss2016}. The linear functional effect and many other additive effects are available using the \textsf{R}~package \pkg{gamlss.add} \citep{gamlss.add2015} to incorporate additive effects of the \textsf{R}~package \pkg{mgcv} \citep{mgcv2016} into the GAMLSS models.

Estimation of GAMLSS by maximizing the penalized log-likelihood using LAML, Section~\ref{sec.mgcv}, is implemented in the \textsf{R}~package \pkg{mgcv} \citep{mgcv2016}, which contains linear functional effects and has many additive terms implemented. Note that in \textsf{R}~package \pkg{mgcv} the Gaussian location scale family models the inverse standard deviation instead of the standard deviation. 

To incorporate functional effects with FPC-basis expansion, any software can be used to estimate the FPCA. The scores can then be used like scalar covariates in the models. We use the \textsf{R}~package \pkg{refund} \citep{refund2016} for the estimation of the FPCA.


\subsection{Example \textsf{R} code}
\label{sec.appendix_code}
The application of the three estimation methods --boosting, gamlss and mgcv-- in \textsf{R} is demonstrated on a small simulated example for a Gaussian location scale model with a linear effect of one scalar and one functional covariate:
\begin{small}
\begin{verbatim}
########### simulate Gaussian data 
library(splines)
n <- 500 ## number of observations
G <- 120 ## number of evaluation points per functional covariate

set.seed(123) ## ensure reproducibility
z <- runif(n) ## scalar covariate
z <- z - mean(z)
s <- seq(0, 1, l = G) ## index of functional covariate
## generate functional covariate
x <- t(replicate(n, drop(bs(s, df = 5, int = TRUE) %*% 
                           runif(5, min = -1, max = 1))))
## center x per observation point
x <- scale(x, center = TRUE, scale = FALSE) 

mu <- 2 + 0.5*z + (1/G*x) %*% sin(s*pi)*5 ## true functions for expectation
sigma <- exp(0.5*z - (1/G*x) %*% cos(s*pi)*2) ## and for standard deviation 
y <- rnorm(mean = mu, sd = sigma, n = n) ## draw respone y_i ~ N(mu_i, sigma_i)

########### fit by boosting
library(gamboostLSS)
library(FDboost)

## save data as list containing s as well
dat_list <- list(y = y, z = z, x = I(x), s = s)
## model fit by boosting 
## bols: linear base-learner for scalar covariates
## bsignal: linear base-learner for functional covariates
m_boost <- FDboostLSS(list(mu = y ~ bols(z, df = 2) + 
                             bsignal(x, s, df = 2, knots = 16), 
                           sigma = y ~ bols(z, df = 2) + 
                             bsignal(x, s, df = 2, knots = 16)), 
                      timeformula = NULL, data = dat_list)
## find optimal number of boosting iterations on a 2D grid in [1, 500]
## using 5-fold bootstrap
grid <-  make.grid(c(mu = 500, sigma = 500), length.out = 10)
## takes some time, easy to parallelize on Linux 
cvr <- cvrisk(m_boost, folds = cv(model.weights(m_boost), B = 5), 
              grid = grid, trace = FALSE)
## use model at optimal stopping iterations 
m_boost <- m_boost[mstop(cvr)] ## m_boost[c(253, 126)] 
summary(m_boost)

## plot smooth effects of functional covariates
par(mfrow = c(1,2))
plot(m_boost$mu, which = 2, ylim = c(0,5))
lines(s, sin(s*pi)*5, col = 3, lwd = 2)
plot(m_boost$sigma, which = 2, ylim = c(-2.5,2.5))
lines(s, -cos(s*pi)*2, col=3, lwd = 2)

## do a QQ-plot of the quantile residuals 
## compute quantile residuals
predmui <- predict(m_boost, parameter = "mu", type = "response")
predsigmai <- predict(m_boost, parameter = "sigma", type = "response")
yi <- m_boost$mu$response
resi_boosting <- (yi - predmui) / predsigmai
#### alternative way to compute the quantile residuals, 
#### which also works for non-Gaussian responses 
## library(gamlss)
## resi_boosting <- qnorm(pNO(q = yi, mu = predmui, sigma = predsigmai))
qqnorm(resi_boosting, main = "", ylim = c(-4, 4), xlim = c(-4, 4))
abline(0, 1, col = "darkgrey", lwd = 0.6, cex = 0.5)


########### fit by gamlss
library(mgcv)
library(gamlss)
library(gamlss.add)

## multiply functional covariate with integration weights
xInt <- 1/G * x 
## sma: matrix containing the evaluation points s in each row 
dat <- data.frame(y = y, z = z, x = I(xInt), 
                  sma = I(t(matrix(s, length(s), n))))
## model fit by gamlss 
## use ga() from gamlss.add to incorporate smooth effects from mgcv 
m_gamlss <- gamlss(y ~ z + 
                     ga( ~ s(sma, by = x, bs = "ps", m = c(2, 1), k = 20)), 
                   sigma.formula = ~ z + 
                     ga( ~ s(sma, by = x, bs = "ps", m = c(2, 1), k = 20)), 
                   data = dat)
summary(m_gamlss)

## plot smooth effects of functional covariates
plot(getSmo(m_gamlss, what = "mu"))
lines(s, sin(s*pi)*5, col = 3, lwd = 2)
plot(getSmo(m_gamlss, what = "sigma"))
lines(s, -cos(s*pi)*2, col = 3, lwd = 2)


########### fit by mgcv
library(mgcv)

## multiply functional covariate with integration weights
xInt <- 1/G * x 
## sma: matrix containing the evaluation points s in each row 
dat <- data.frame(y = y, z = z, x = I(xInt), 
                  sma = I(t(matrix(s, length(s), n))))

## model fit by mgcv 
## note that gaulss in mgcv models the inverse standard deviation  
m_mgcv <- gam(list(y ~ z + 
                     s(sma, by = x, bs = "ps", m = c(2, 1), k = 20), 
                    ~ z +
                     s(sma, by = x, bs = "ps", m = c(2, 1), k = 20)), 
                   data = dat, family = gaulss)
summary(m_mgcv)

## plot smooth effects of functional covariates
plot(m_mgcv, select = 1)
lines(s, sin(s*pi)*5, col = 3, lwd = 2)
plot(m_mgcv, select = 2)
lines(s, -cos(s*pi)*2, col = 3, lwd = 2)


########### fit with FPC-basis for the example of mgcv
library(refund)

## do the FPCA on x; explain 99% of variability
kl_x <- fpca.sc(x, pve = 0.99, argvals = s)
## scores that can be used like scalar covariates
scores <- kl_x$scores
dat <- data.frame(y = y, z = z, scores = scores)

## model fit by mgcv
m_mgcv_fpca <- gam(list(y ~ z + scores, 
                        ~ z + scores), 
                   data = dat, family = gaulss)
summary(m_mgcv_fpca)

## plot smooth effects of functional covariates
coefs_scores <- coef(m_mgcv_fpca)[
  grepl("scores", names(coef(m_mgcv_fpca)))]
coefs_scores_mu <- coefs_scores[1:kl_x$npc]
coefs_scores_sigma <- coefs_scores[(kl_x$npc+1):length(coefs_scores)]
plot(s, kl_x$efunctions %*% coefs_scores_mu, type = "l")
lines(s, sin(s*pi)*5, col = 3, lwd = 2)
plot(s, kl_x$efunctions %*% coefs_scores_sigma, type = "l")
lines(s, -cos(s*pi)*2, col = 3, lwd = 2)
\end{verbatim}
\end{small}


\section{Further results of the application on stock returns}
\label{sec.appendix_stocks}
\begin{figure}[ht]\centering
\makebox{
\raisebox{0.2\textwidth}{sc\phantom{+fun }}
\includegraphics[width=0.34\textwidth, page=3]{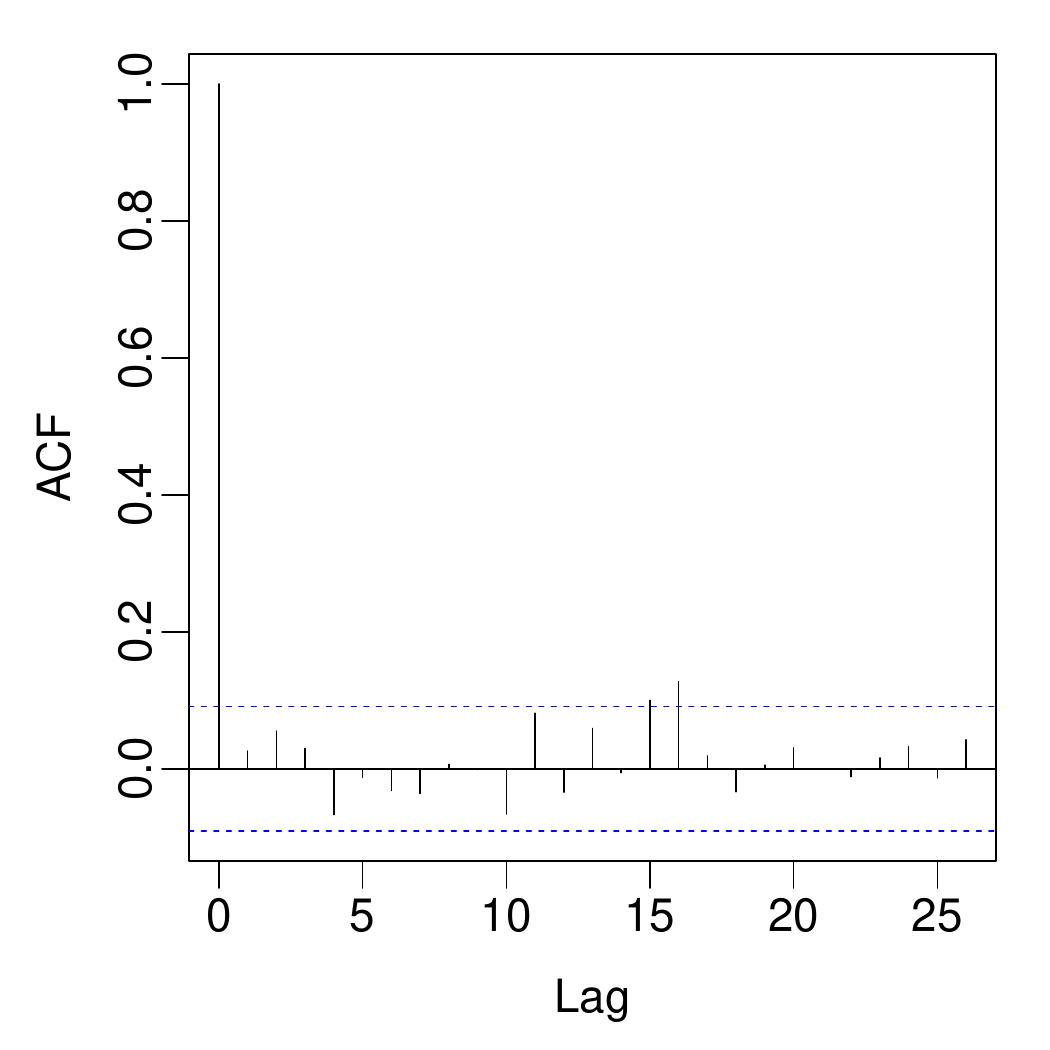}
\includegraphics[width=0.34\textwidth, page=3]{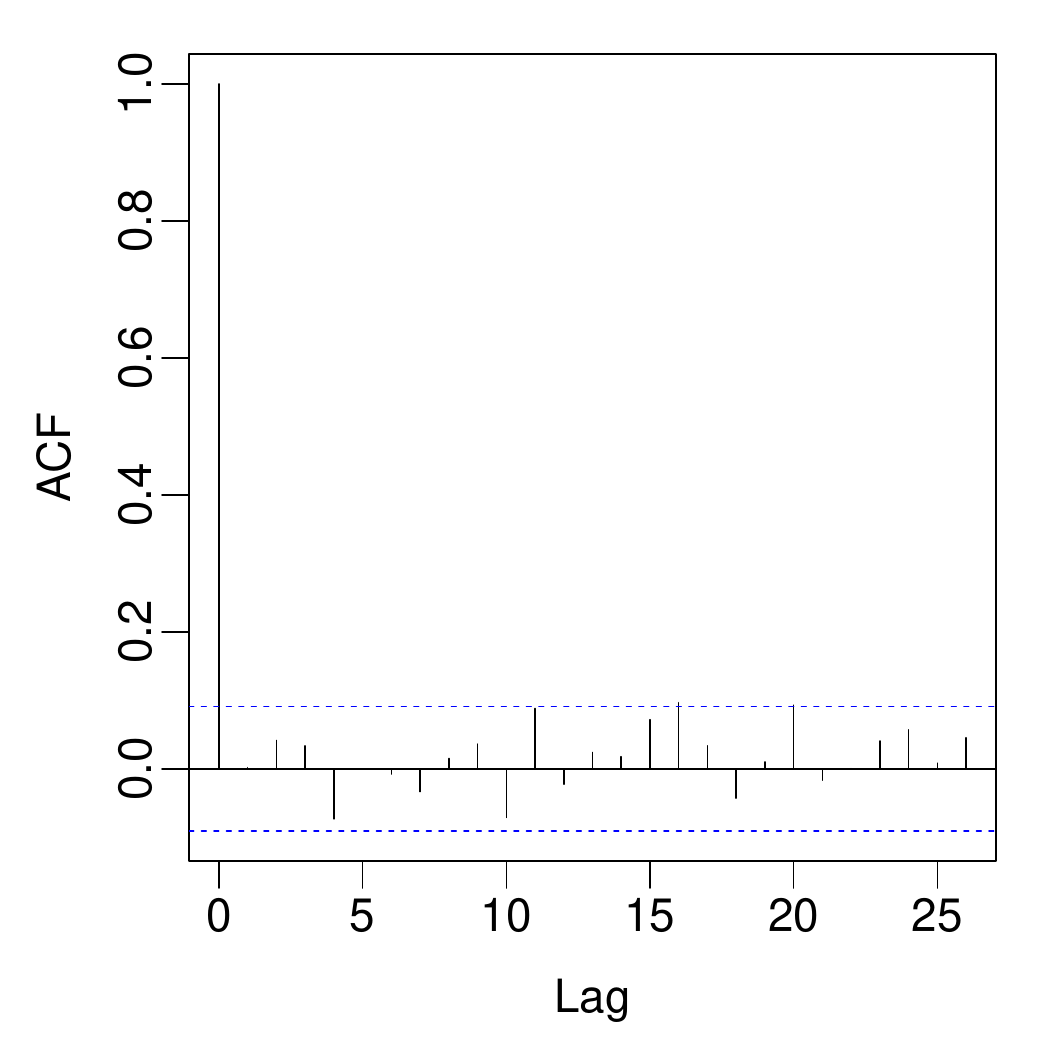}
}
\makebox{
\raisebox{0.2\textwidth}{sc+fun }
\includegraphics[width=0.34\textwidth, page=3]{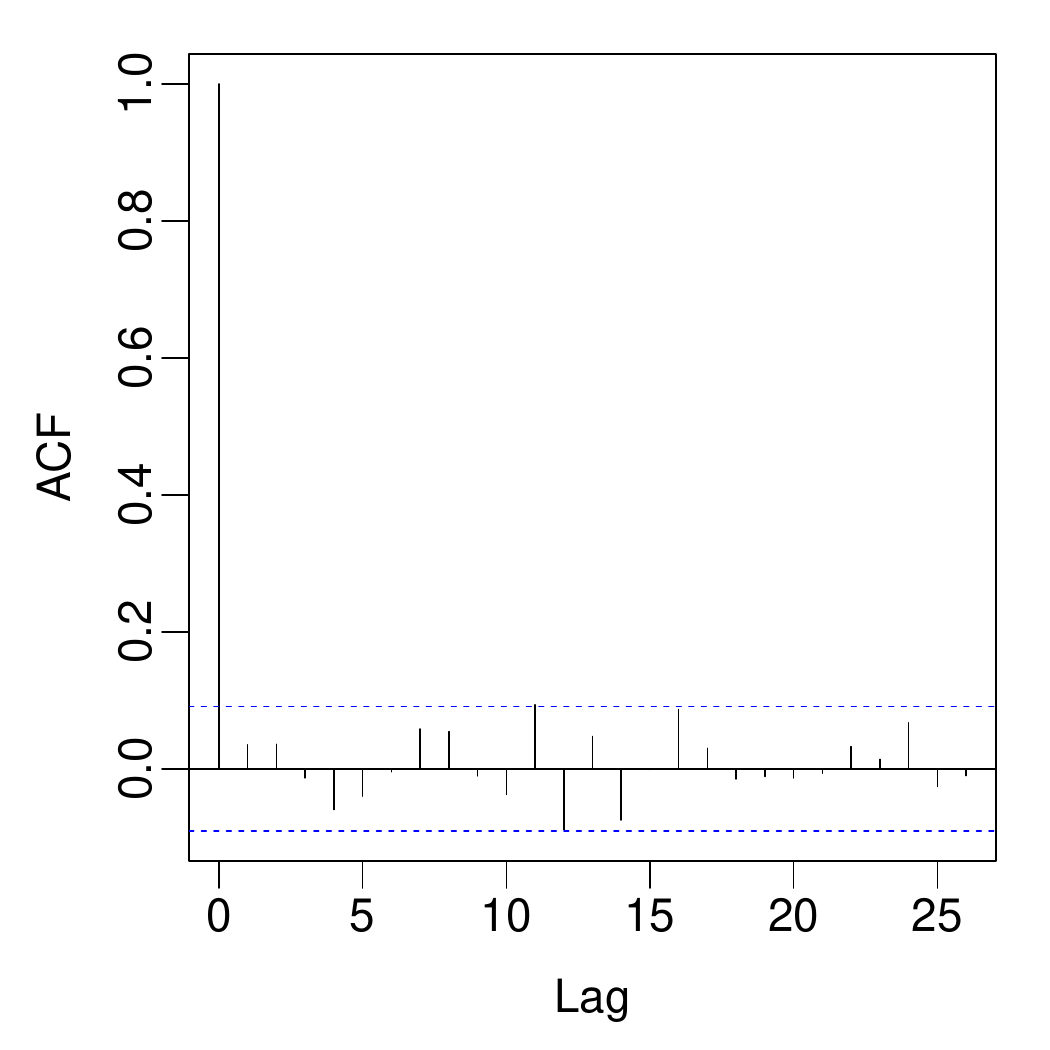}
\includegraphics[width=0.34\textwidth, page=3]{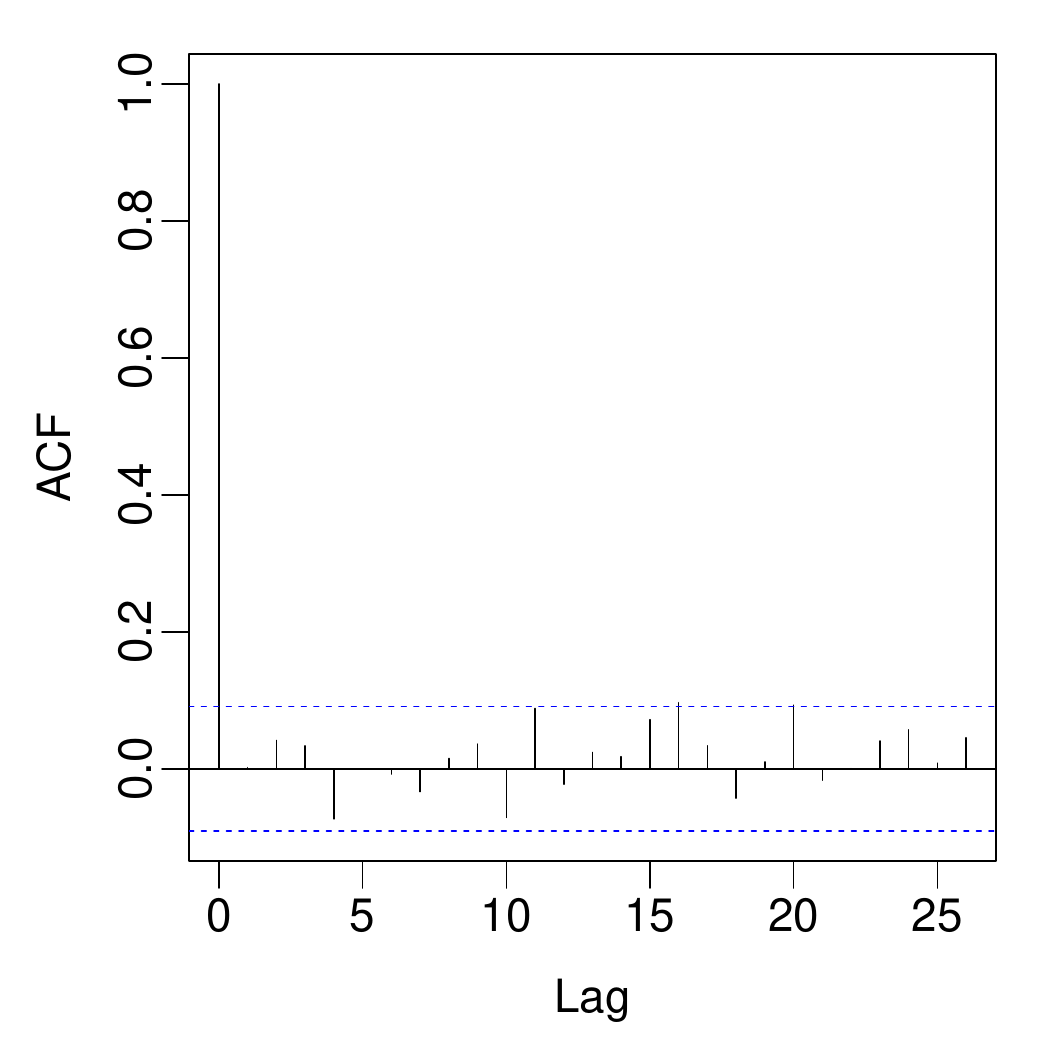}
}
\caption{Model choice for the application: The estimated ACF of the squared quantile residuals in the GAMLSS with scalar variables (sc, top row) or with scalar and functional variables (sc+fun, bottom row) assuming normally distributed response (left column) or Students $t$-distribution for the response (right column) for models estimated by the gamlss algorithm. The dashed blue line marks the limits for point-wise 95\% tests for the ACF to be zero. }
\label{fig.acf}
\end{figure}

\begin{figure}[ht]\centering
 \includegraphics[width=0.3\textwidth, page=1]{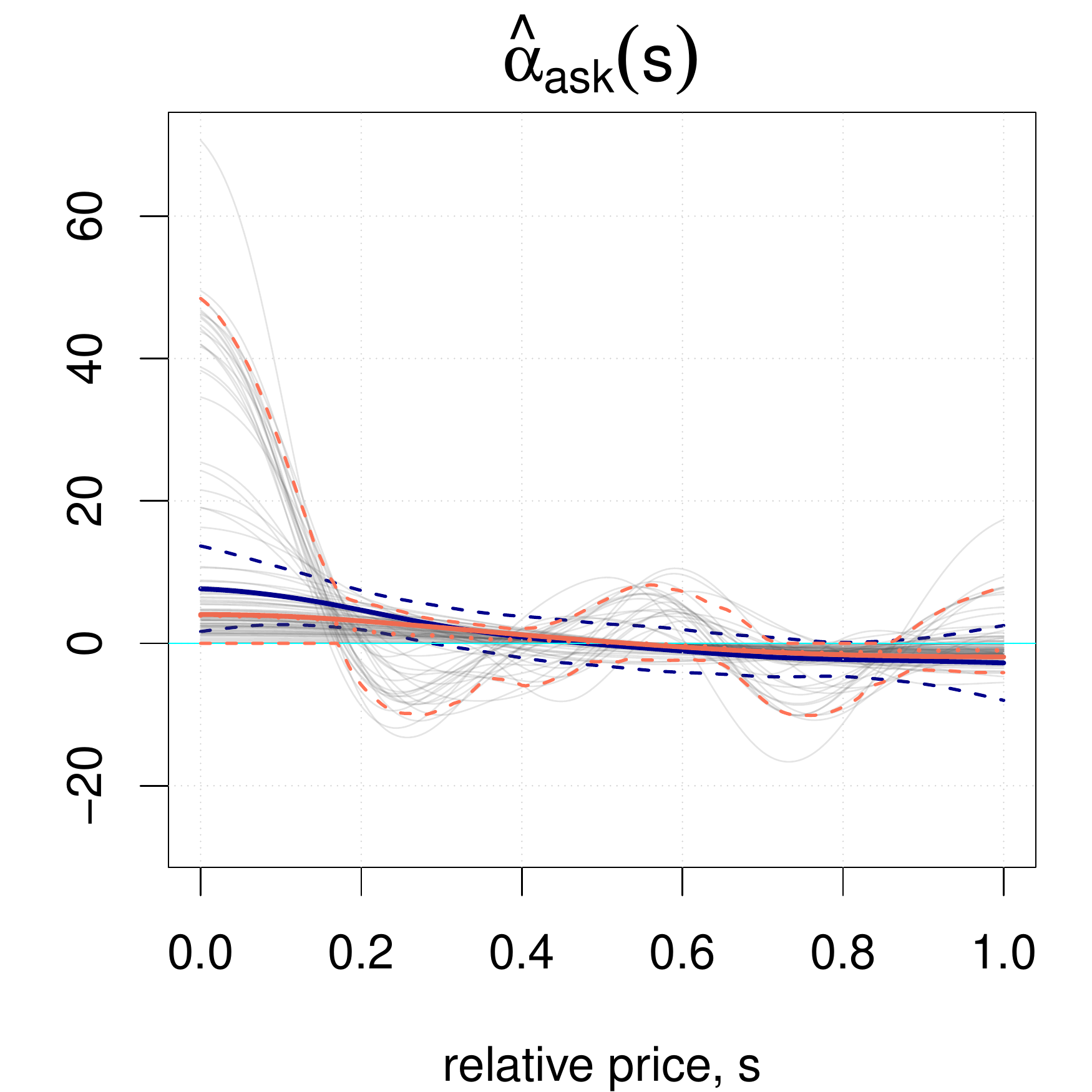}
 \includegraphics[width=0.3\textwidth, page=2]{Gaussian_boost_mgcv_90_shrink10_intercept_coefBS.pdf}
 \includegraphics[width=0.3\textwidth, page=3]{Gaussian_boost_mgcv_90_shrink10_intercept_coefBS.pdf}
 \includegraphics[width=0.3\textwidth, page=4]{Gaussian_boost_mgcv_90_shrink10_intercept_coefBS.pdf}
 \includegraphics[width=0.3\textwidth, page=5]{Gaussian_boost_mgcv_90_shrink10_intercept_coefBS.pdf}
 \includegraphics[width=0.3\textwidth, page=6]{Gaussian_boost_mgcv_90_shrink10_intercept_coefBS.pdf}
\caption{Results for the application, model with normal distribution: Estimated coefficients for $\mu_i$ (top panel) and $\sigma_i$ (bottom panel) in the GAMLSS with the two liquidities as functional covariates and $p_1=p_2=10$ lag variables.
For the intercept of the standard deviation we plot $\hat{\beta}_0 - 1$ to better fit the intercept into the range of the lag effects.
The boosting estimates on the 100 block-bootstrap samples are plotted as partly transparent lines or circles and the point-wise 2.5, 50, and 97.5\% quantiles as dashed orange lines. The boosting estimates are plotted as solid orange line.
The estimates of mgcv using ten P-splines with shrinkage penalty with point-wise 95\% confidence bands are plotted in dark blue. The zero-line is marked with a light-blue line.}
\label{fig.coef_shrink10}
\end{figure}

\begin{figure}[ht]\centering
 \includegraphics[width=0.3\textwidth, page=1]{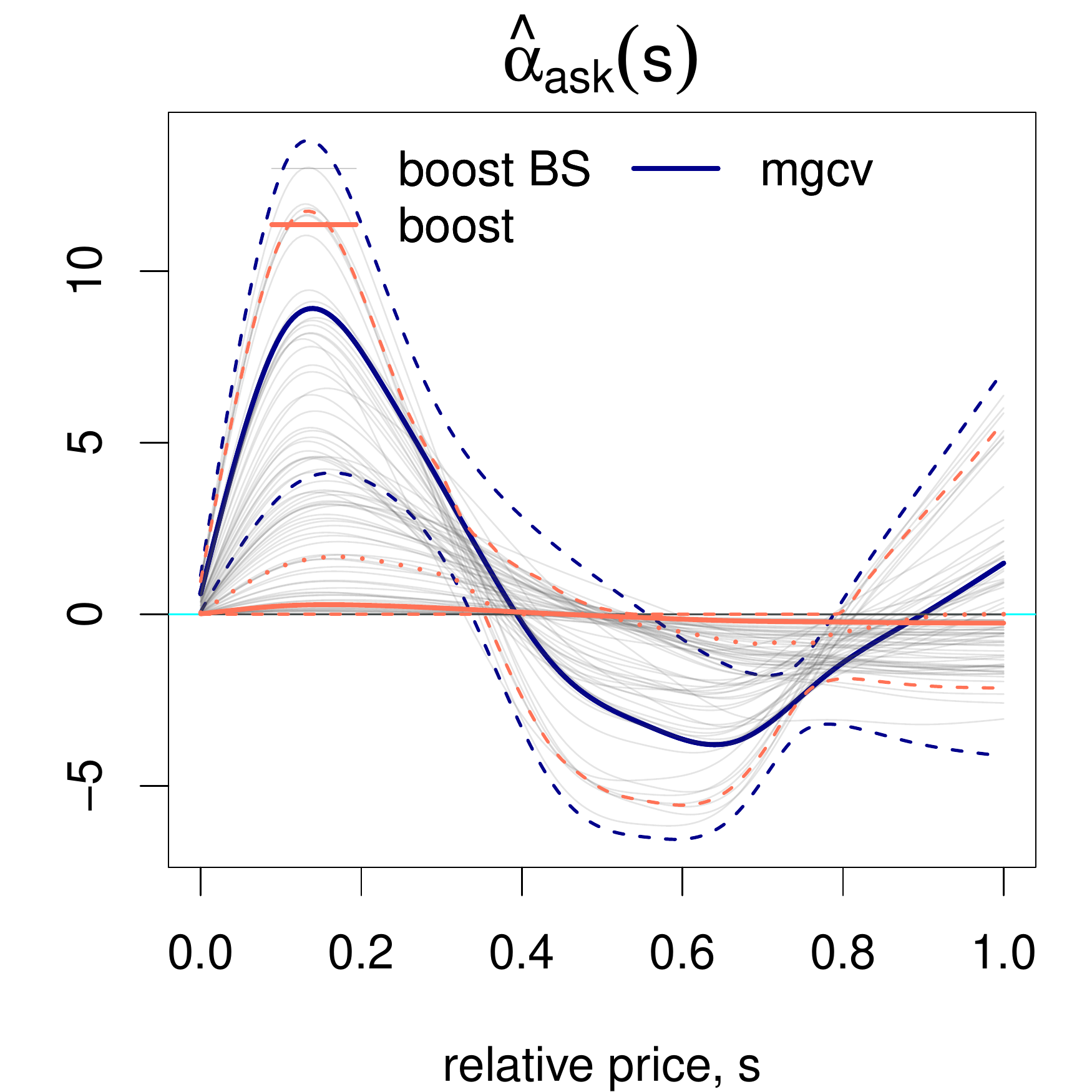}
 \includegraphics[width=0.3\textwidth, page=2]{Gaussian_scfunFPCA_90_boost_mgcv_coefBS.pdf}
 \includegraphics[width=0.3\textwidth, page=3]{Gaussian_scfunFPCA_90_boost_mgcv_coefBS.pdf}
 \includegraphics[width=0.3\textwidth, page=4]{Gaussian_scfunFPCA_90_boost_mgcv_coefBS.pdf}
 \includegraphics[width=0.3\textwidth, page=5]{Gaussian_scfunFPCA_90_boost_mgcv_coefBS.pdf}
 \includegraphics[width=0.3\textwidth, page=6]{Gaussian_scfunFPCA_90_boost_mgcv_coefBS.pdf}
\caption{Results for the application, model with normal distribution using FPC basis functions: Estimated coefficients for $\mu_i$ (top panel) and $\sigma_i$ (bottom panel) in the GAMLSS with the two liquidities as functional covariates and $p_1=p_2=10$ lag variables.
For the intercept of the standard deviation we plot $\hat{\beta}_0 - 1$ to better fit the intercept into the range of the lag effects.
The boosting estimates on the 100 block-bootstrap samples are plotted as partly transparent lines or circles and the point-wise 2.5, 50, and 97.5\% quantiles as dashed orange lines. The boosting estimates are plotted as solid orange line.
The estimates of mgcv with point-wise 95\% confidence bands are plotted in dark blue. The zero-line is marked with a light-blue line.}
\label{fig.coef_FPCA}
\end{figure}

\begin{figure}[hp]\centering
 \includegraphics[width=0.3\textwidth, page=1]{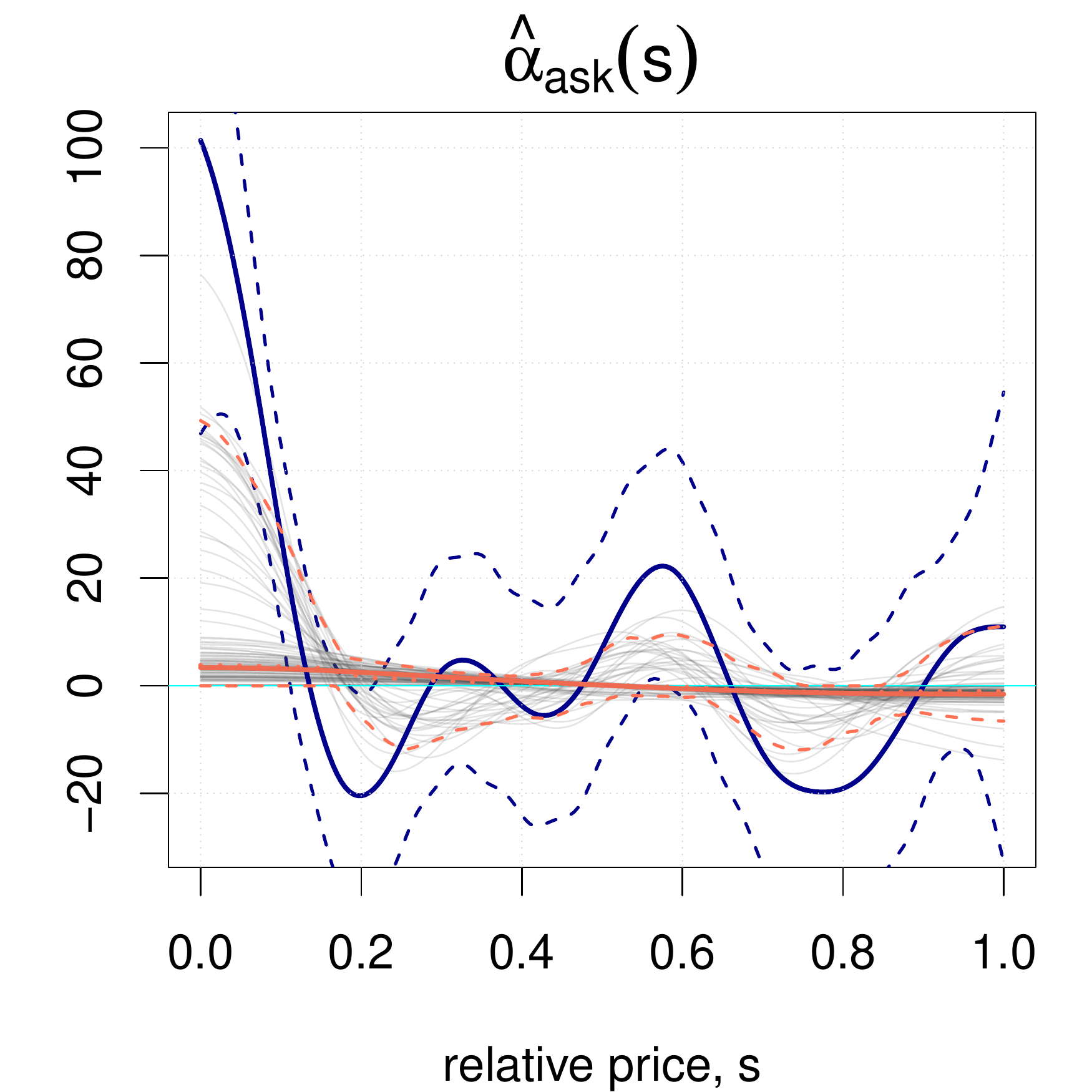}
 \includegraphics[width=0.3\textwidth, page=2]{student_boost_gamlss_90_intercept_coefBS.pdf}
 \includegraphics[width=0.3\textwidth, page=3]{student_boost_gamlss_90_intercept_coefBS.pdf}
 \includegraphics[width=0.3\textwidth, page=4]{student_boost_gamlss_90_intercept_coefBS.pdf}
 \includegraphics[width=0.3\textwidth, page=5]{student_boost_gamlss_90_intercept_coefBS.pdf}
 \includegraphics[width=0.3\textwidth, page=6]{student_boost_gamlss_90_intercept_coefBS.pdf}
  \includegraphics[width=0.3\textwidth, page=7]{student_boost_gamlss_90_intercept_coefBS.pdf}
\caption{Results for the application, model with $t$-distribution: Estimated coefficients for $\mu_i$ (top), $\sigma_i$ (middle) and df (bottom) in the GAMLSS with the two liquidities as functional covariates and $p_1=p_2=10$ lag variables.
For the intercept of the standard deviation we plot $\hat{\beta}_0 - 1$ to better fit the intercept into the range of the lag effects.
The boosting estimates on the 100 block-bootstrap samples are plotted as partly transparent lines or circles and the point-wise 2.5, 50, and 97.5\% quantiles as dashed orange lines. The boosting estimates are plotted as solid orange line.
The estimates of gamlss with point-wise 95\% confidence bands are plotted in dark blue. The zero-line is marked with a light-blue line.}
\label{fig.coef_t}
\end{figure}

In Figure~\ref{fig.acf} the estimated ACF is plotted for the squared residuals of models fitted by gamlss.
In Figure~\ref{fig.coef_shrink10} the estimates for the model assuming a normal location scale model are plotted for boosting, as in Figure~\ref{fig.coef},  
and for mgcv the estimates resulting from using the shrinkage-penalty of \cite{marra2011} with $K_j=10$ basis functions are given.
In Figure~\ref{fig.coef_FPCA} the estimates for the model assuming a normal location scale model are plotted for boosting and for mgcv using FPC basis functions to represent both,  the functional covariate and the functional coefficient. We use the first 3 eigenfunctions as those explain more than 99\% of the variability in the bid- and ask-curves. The confidence intervals are computed conditional on the estimated eigendecomposition. 
In Figure~\ref{fig.coef_t} the estimates for the model assuming $t$-distributed response are plotted for boosting and gamlss. The estimates for $\mu$ and $\sigma$ are very similar to those for a model assuming normal distribution, compare Figure~\ref{fig.coef}. 
The estimated df for gamlss are $\exp(\hat{\gamma}_0) \approx 8.2$ and for boosting $\exp(\hat{\gamma}_0) \approx 3.8$.
%

\section{Details of the general simulation study}
\label{sec.sim_appendix}
\begin{figure}[ht]\centering
  \includegraphics[width=0.3\textwidth, page=1]{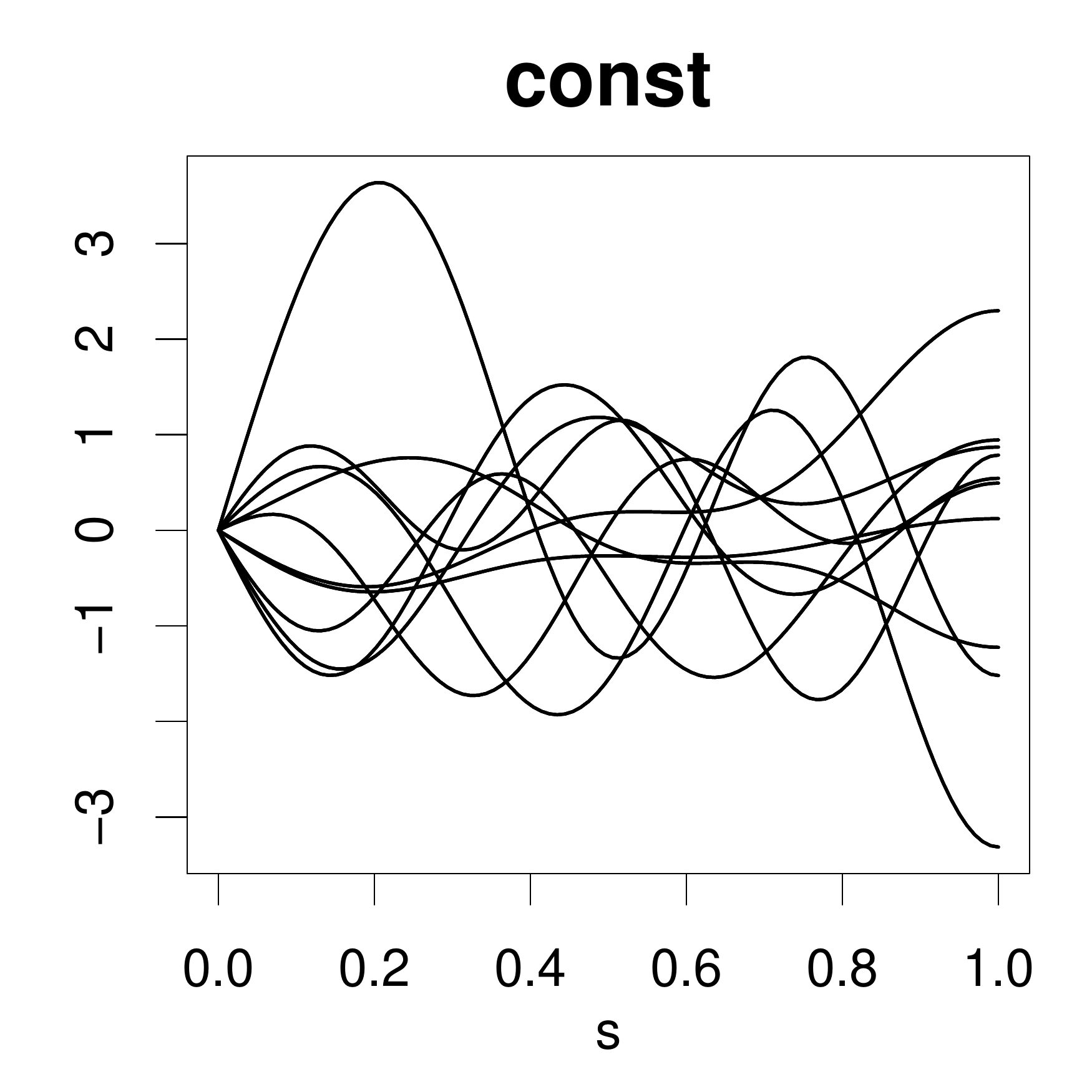}
  \includegraphics[width=0.3\textwidth, page=3]{dataSimPaper.pdf}
  \includegraphics[width=0.3\textwidth, page=5]{dataSimPaper.pdf}
  \includegraphics[width=0.3\textwidth, page=2]{dataSimPaper.pdf}
  \includegraphics[width=0.3\textwidth, page=4]{dataSimPaper.pdf}
  \includegraphics[width=0.3\textwidth, page=6]{dataSimPaper.pdf}
\caption{Draws from simulated data settings for the functional covariates;
         top row: start in zero, bottom row: start with random point; 
         from left to right: functional variables simulated using random coefficients with constant,
         linearly decreasing and
         exponentially decreasing variances.}
\label{fig.simcov}
\end{figure}

\begin{figure}[ht]\centering
  \includegraphics[width=0.9\textwidth, page=1]{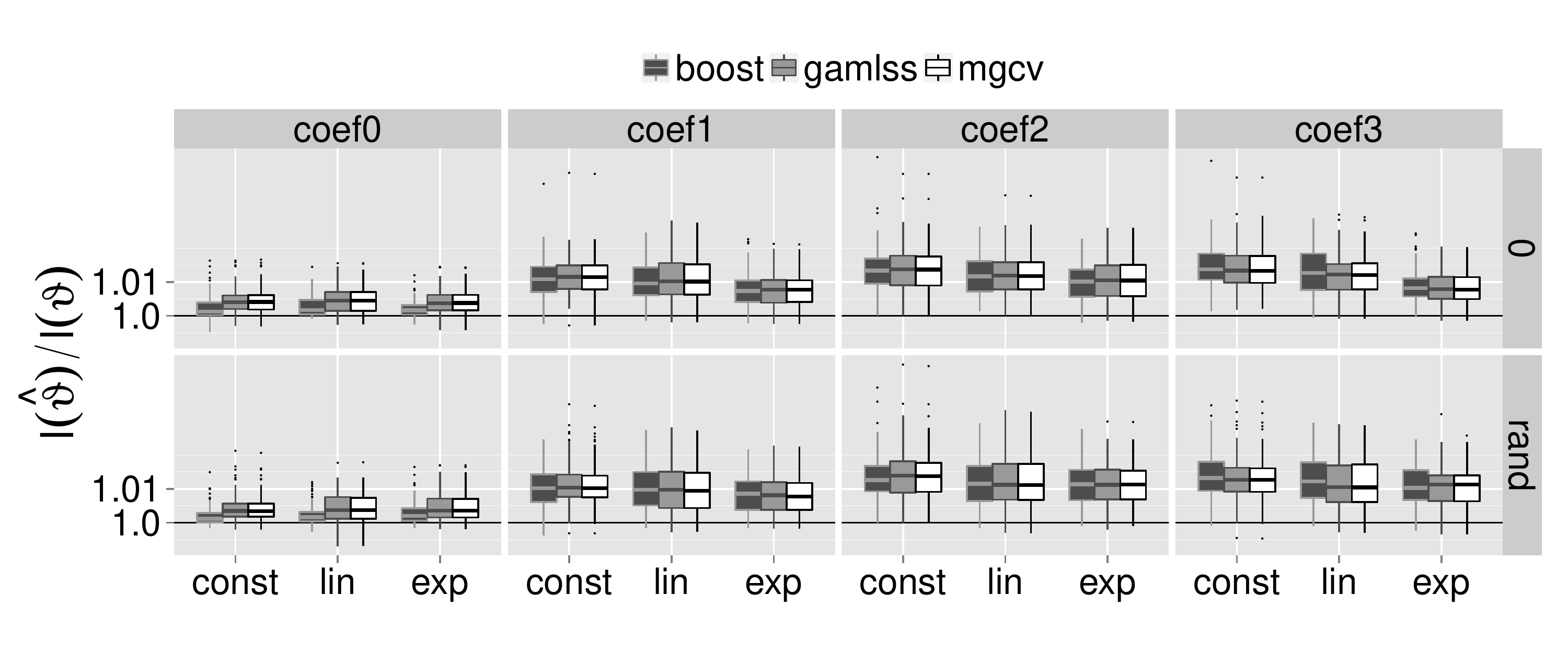}
\caption{Results from the simulation study: The quotient of the negative log-likelihoods for the predicted and the true distribution parameters in the different data settings with constant, linearly and exponentially decreasing variances for the random coefficients generating the functional covariates on the x-axis and functions starting in zero or in a random point in rows. The different functional coefficients are given in columns from coef0 to coef3. The three fitting algorithms are color coded. The one-line is marked as in this case the likelihood of the predicted and the true parameters is equal. Note that all values are displayed on a logarithmic scale.}
\label{fig.simlike}
\end{figure}

\begin{figure}[ht]\centering
\makebox{
\raisebox{0.08\textwidth}{exp\phantom{st, rand}}
  \includegraphics[width=0.44\textwidth, page=1]{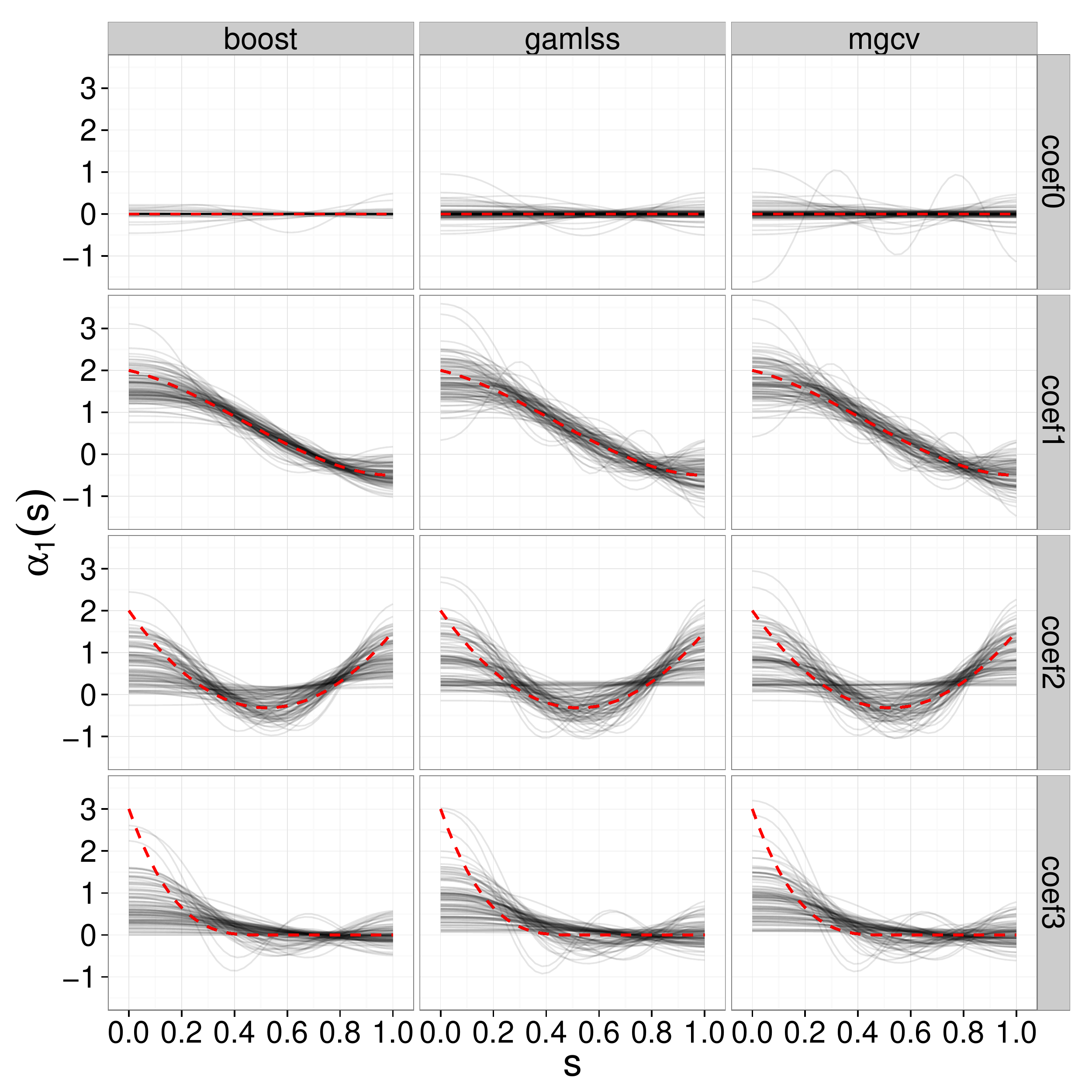}
  \includegraphics[width=0.44\textwidth, page=1]{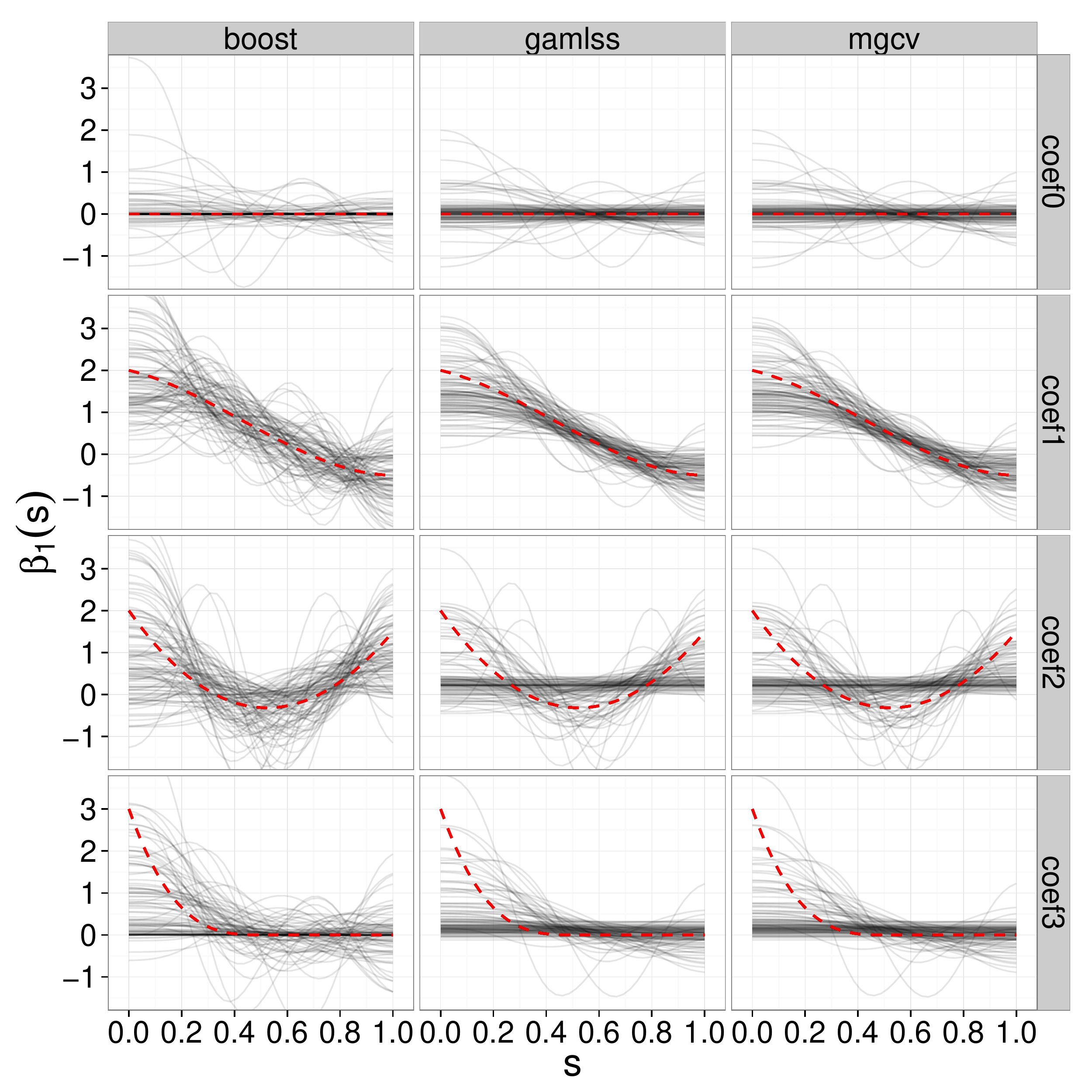}
  }
\makebox{
\raisebox{0.08\textwidth}{const, rand}
  \includegraphics[width=0.44\textwidth, page=1]{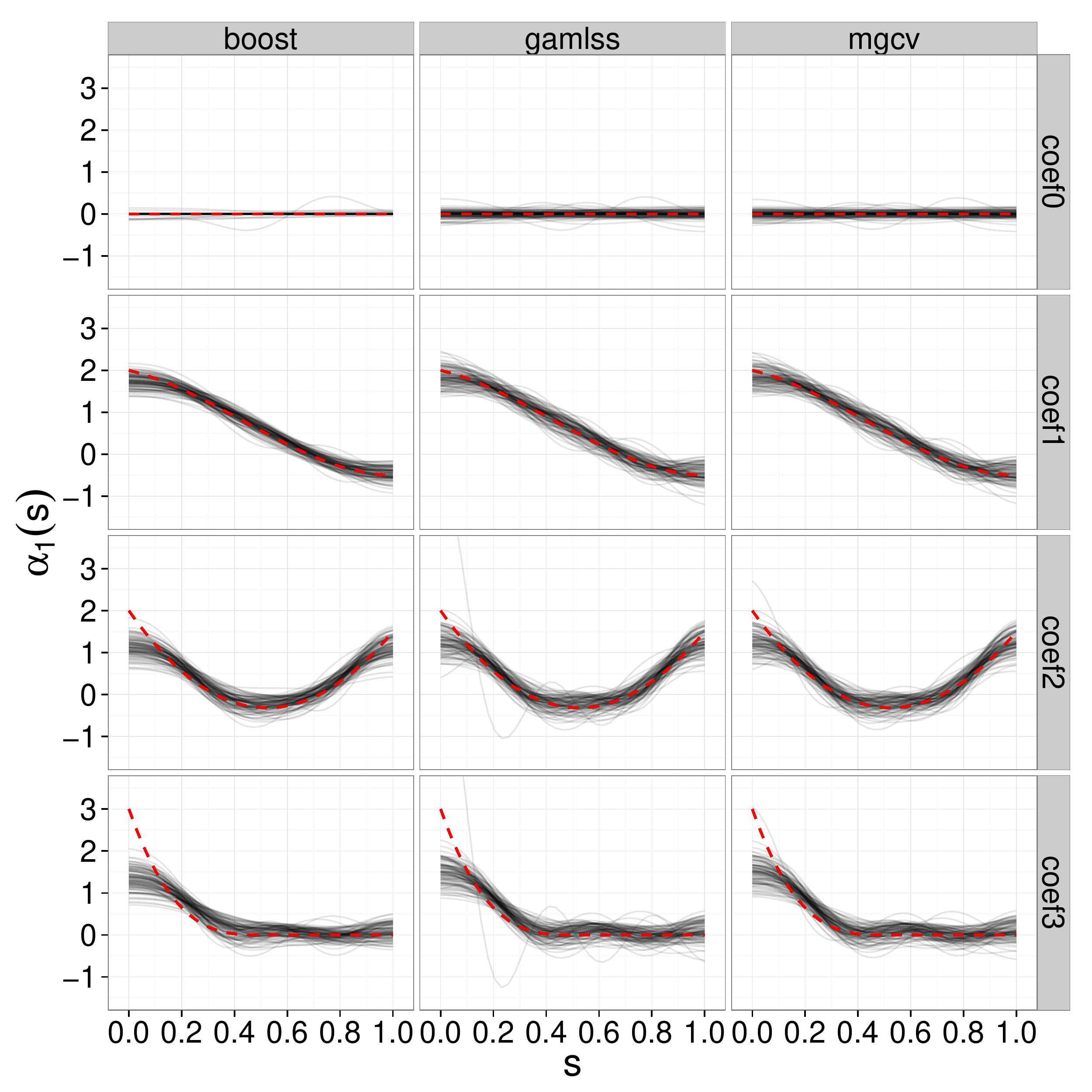}
  \includegraphics[width=0.44\textwidth, page=1]{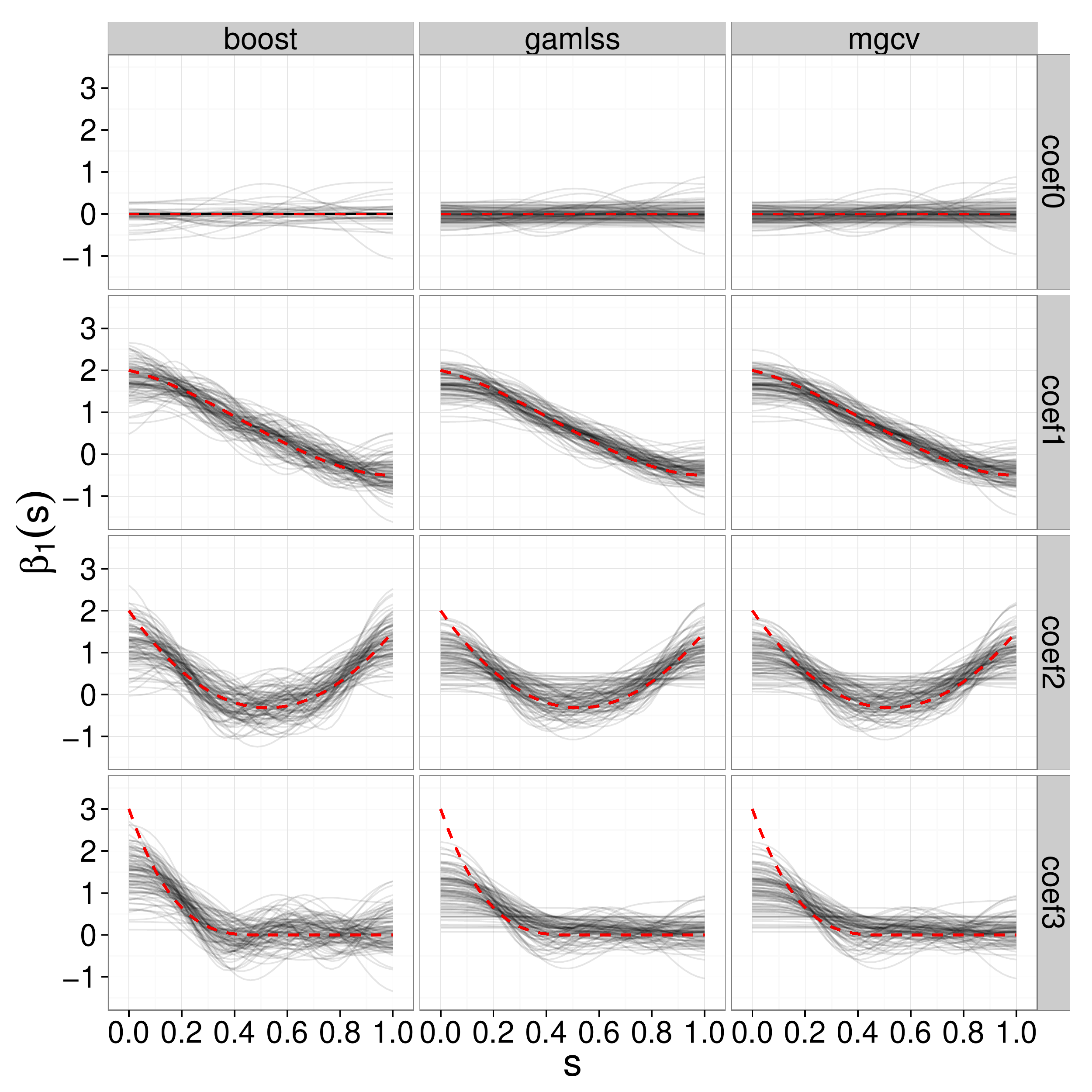}
}
\caption{Results from the simulation study: Estimated coefficient functions in the 100 runs per setting for the three fitting algorithms; functional covariates with exponentially decreasing variances and starting point zero (top) or constant variances   with the addition of a random variable to get random starting points (bottom); estimates for the expectation (left) and the standard deviation (right). The true coefficient functions are given as bold dashed lines.}
\label{fig.simhatcoef}
\end{figure}

In this simulation study, the model fits of the three implementations are compared systematically for data situations with different complexities, see Section~\ref{sec.genSim} for a summary of the results. 
\\\\
\underline{Data generation.}
For the simulation study we consider a model with normally distributed response, where the expectation and the standard deviation of the response $y_i$ depend on two functional covariates $x_{i1}(s)$, $x_{i2}(s)$, with $s\in[0,1]$, $i=1, \ldots, N$, following the model:
\begin{equation*}
\label{eq.simModel}
\begin{split}
y_i| x_i  \sim N(\mu_i,\ \sigma_i^2),  \\
\mu_i = h^{(\mu)}(x_i) = \alpha_0 + \sum_{j=1}^{2} \int x_{ij}(s) \alpha_j(s)\, \mathrm{d}s, \\
\log \sigma_i = h^{(\sigma)}(x_i) =\beta_0 + \sum_{j=1}^{2} \int x_{ij}(s) \beta_j(s)\, \mathrm{d}s.
\end{split}
\end{equation*}
For some simulation settings the coefficient functions are completely 0, inducing non-informativeness of the variable for the corresponding parameter.
We draw $N=500$ observations for each combination of the following settings from the normal location scale model:
\begin{enumerate}
\item Simulate functional covariates $x_j(s)$, with 100 equally spaced evaluation points $(s_{1}, \ldots,s_{100})^\top$, using $C=5$ 
 basis functions $\phi_{c}(s) = \sqrt{2}\sin(\pi(c - 0.5)s)$, $c = 1, \ldots, C$, with random coefficients from a $C$-dimensional normal with $N_C \left(\boldsymbol{0}, \diag(\zeta_1, \ldots, \zeta_C) \right)$, and
\begin{description}
\item[\textit{const}] variances $\zeta_{c} = 1 $, yielding constant variances.
\item[\textit{lin}] variances $\zeta_{c}= 0.1 c$, yielding linearly decreasing variances.
\item[\textit{exp}] variances $\zeta_{c}=[\pi(c-0.5)]^{-2}$, yielding exponentially decreasing variances. Using the Karhunen-Lo\`eve expansion with orthogonal basis functions $\phi_{c}(s)$, weighted with normally distributed random variables $N_C \left(\boldsymbol{0}, \diag(\zeta_1,\ldots, \zeta_C) \right)$, this would yield draws from a Wiener process for $C \to \infty$. 
\end{description}
All those settings generate functional variables starting in zero. Additionally, we consider settings that start in a random point:
\begin{description}
\item[\textit{rand}] simulate functional variables as above and add a $N(0,\zeta_0)$ random variable, with $\zeta_0 = \zeta_1$ for each setting.
\end{description} 
The data generation is constructed such that the covariates carry different amounts of information. The covariates carry most information in the setting with constant variances, less for linearly decreasing and least for exponentially deceasing variances \citep{scheipl2016}.   
The covariates are centered for each evaluation point to induce a mean effect of zero, i.e., $\sum_i \int \beta_j(s) x_{ij}(s)\, \mathrm{d}s = 0$. Then the covariates are standardized with their global empirical standard deviation to make the effect size comparable over all settings. 
\item The coefficient functions $\alpha_j(s)$, $\beta_j(s)$ to model the expectation and the standard deviation are
\begin{description}
\item[\textit{coef0}] completely zero.
\item[\textit{coef1}] four cubic B-splines with coefficients $(2, 1.5, -0.5, -0.5)$, giving a decreasing curve. 
\item[\textit{coef2}] four cubic B-splines with coefficients $(0.5, -1, -1, 1.5)$, giving a u-shaped curve. 
\item[\textit{coef3}] four cubic B-splines with coefficients $(3, 0, 0, 0)$, giving a curve that is quite high for $s=0$ with a steep decrease.
\end{description} 
\end{enumerate}
We run 100 replications for each data generation combination. In Figure~\ref{fig.simcov} ten simulated observations per data-generating process are depicted. The true coefficient functions can be seen in Figure~\ref{fig.simhatcoef}.
\\\\
\underline{Estimation.}
For the estimation of the models we specify normal location scale models and use one of the three estimation algorithms boosting \citep{mayr2012,brockhaus2015}, gamlss \citep{rigby2005,rigby2014} and mgcv \citep{wood2011,wood2015}. 
The smooth effects are estimated using 20 cubic P-splines with first order difference penalties.
For the boosting algorithm, the step-length $\nu=(0.1, 0.01)^\top$ is fixed and the optimal stopping iterations are searched on a two-dimensional grid containing values from 1 to 5000. The smoothing parameters $\lambda_j^{(q)}$ are chosen such that the degrees of freedom are two per base-learner.
For the likelihood-based methods the smoothing parameters $\lambda_j^{(q)}$ are estimated using a REML-criterion.
\\\\
\underline{Simulation results.}
As test data we generate a dataset with 500 observations to evaluate the model predictions out-of-bag. 
To evaluate the goodness of prediction of the model we compute the quotient of the log-likelihood with predicted distribution parameters and the log-likelihood with the true parameters:
\[
 \frac{\sum_{i=1}^{N_{\text{test}}} l(\hat{\mvartheta}_i, y_i)}{ \sum_{i=1}^{N_{\text{test}}} l(\mvartheta_i, y_i)},
\]
where the $y_i$ are the $N_{\text{test}}=500$ response observations in the test data, $\hat{\mvartheta}_i$ are the predictions of the distribution parameters given the model and $\mvartheta_i$ are the true distribution parameters.
To evaluate the accuracy of the estimation of the coefficient functions, we compute the MSE integrated over the domain of the functional covariate, e.g.~for the coefficient~$\beta_j$
\[
\text{MSE}(\beta_j) = \int \left[\beta_j(s) - \hat{\beta}_j(s) \right]^2 \, \mathrm{d}s .
\] 
In Figure~\ref{fig.simlike} the quotient of the log-likelihoods with the predicted and the true parameters is depicted for the different fitting algorithms, grouped by the complexity of the linear predictors -- constant, linearly or exponentially decreasing variances on the x-axis and the addition of a random starting value in the bottom row.
For the considered settings all algorithms yield similar values. In the case of zero-coefficients, boosting seems to outperform the other two methods. The quotient generally becomes higher for more complex linear predictors.

\begin{figure}[ht]\centering
  \includegraphics[width=0.9\textwidth, page=1]{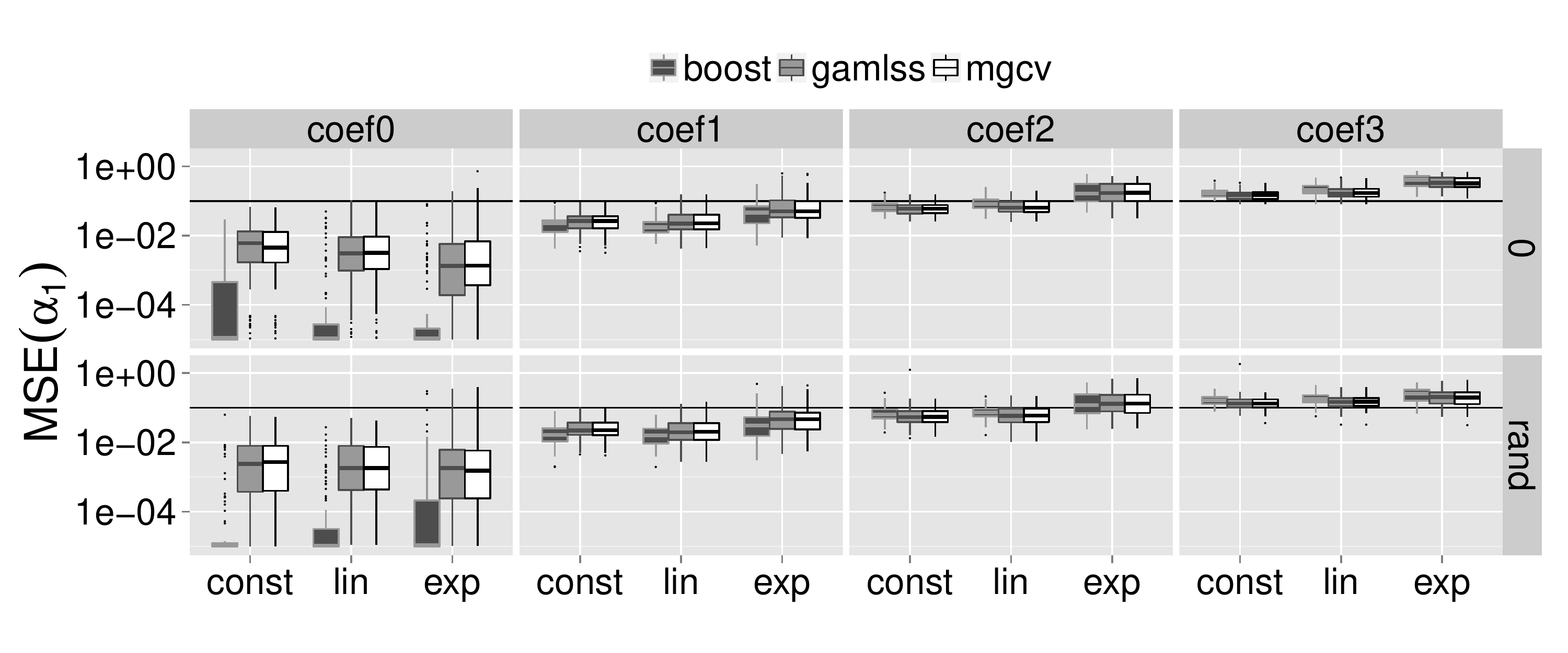}
  \includegraphics[width=0.9\textwidth, page=1]{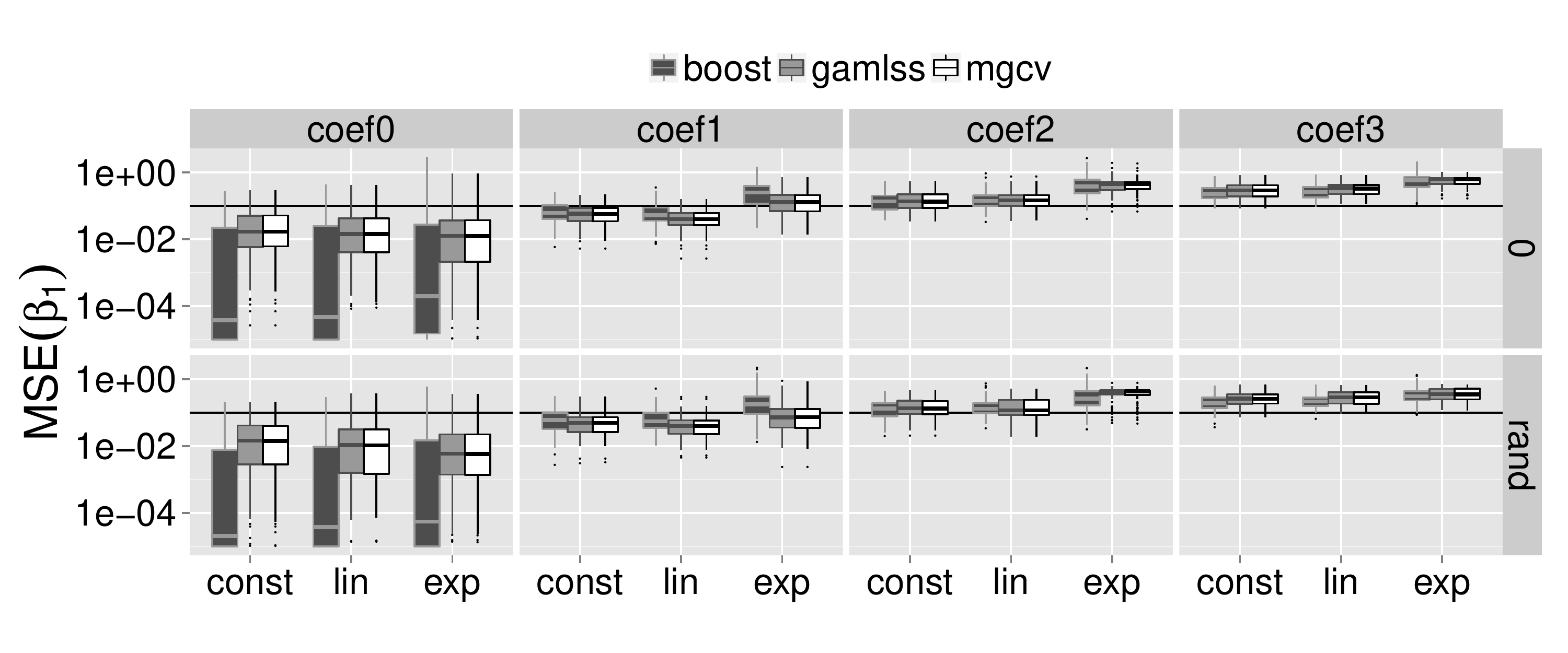}
\caption{Results from the simulation study: MSE of the estimated coefficient functions in the linear predictor of the expectation (top) and of the standard deviation (bottom) for the different data generating settings and fitting algorithms. The three fitting algorithms are color coded. Note that all values are displayed on a logarithmic scale.}
\label{fig.simsd1}
\end{figure}

In Figure~\ref{fig.simsd1} the MSE of the coefficient functions $\alpha_1(s)$ and $\beta_1(s)$ for the expectation and the standard deviation, are plotted grouped by the true coefficient function, data generating process and fitting algorithm. The horizontal 0.1-line is marked, as an MSE smaller 0.1 usually means that the estimated coefficient function is quite similar to the true one.
It can be seen that for equal settings the functional coefficients in the linear predictor of the expectation are fitted with smaller MSE than those of the standard deviation.
Boosting fits better in the case of zero-coefficients, as it conducts model-selection during estimation. For the other settings, the three estimation methods yield similar results. Generally, the MSE is smaller for less complex coefficient functions. Comparing the different data-generating processes, the MSE is lowest for constant variances and highest for exponentially decreasing variances, with the linearly decreasing setting in between, as those settings generate functional variables with decreasing information contained in the functional covariates.
The only exception is the setting with zero-functional coefficients where all data settings have similar MSE. 
There is no big difference between the settings with random starting point and zero as starting point, which means that the estimation is generally not worse for functional covariates that all start in zero.
The highest MSEs are obtained for the functional coefficient coef3 which is high in the beginning and then very steep (Figure~\ref{fig.simsd1}, far right).

The estimated coefficient functions together with the true coefficient functions are plotted in Figure~\ref{fig.simhatcoef} for two data-settings: functional covariates with exponentially decreasing variances and starting point zero (top row) and functional covariates with constant variances and random starting point.
The coefficient functions in the linear predictor of the expectation are estimated more adequately than those of the standard deviation. The estimates are closer to the truth for the more informative data-setting, i.e.~they are better for constant than for exponentially decreasing variances. For the difficult settings, that is decreasing variance in the generation of the functional covariates, and coef2 or coef3, it can be seen that the coefficient function is sometimes estimated as a constant line close to or exactly zero. 

The functional covariates in the application on stock returns are best reflected by the setting with exponentially decreasing variances and starting point zero, which is the most difficult data setting. In this case the absolute value of the coefficients is often underestimated.   
For settings with more informative functional covariates the estimates are mostly close to the true coefficient functions and the three estimation approaches yield very similar estimates and predictions.

\end{appendix}


\bibliography{literatur_entries}  
\bibliographystyle{chicago}  

\end{document}